\documentclass[12pt,preprint]{aastex}

\usepackage{epsfig}
\def\chan{\it Chandra}

\def\hstwfpc2{\it HST/WFPC2}
\def\hst{\it HST}
\def\hstacs{\it HST/ACS}
\def\s3{$S3$}
\def\d25{$d_{25}$}
\def\D25{$D_{25}$}
\def\psf{point spread function}

\shorttitle{LMXBs in 6 Ellipticals}
\shortauthors{Kim et al.}

\begin{document}

\title{Low Mass X-ray Binaries in 6 Elliptical Galaxies:\\
       Connection to Globular Clusters}

\author{Eunhyeuk Kim\altaffilmark{1}, Dong-Woo Kim\altaffilmark{1},
        Giuseppina Fabbiano\altaffilmark{1},
        Myung Gyoon Lee\altaffilmark{2},\\
        Hong Soo Park\altaffilmark{2},
        Doug Geisler\altaffilmark{3} and Boris Dirsch\altaffilmark{3} }

\altaffiltext{1}{Harvard-Smithsonian Center for Astrophysics,
 60 Garden St., Cambridge, MA, 02138}
\altaffiltext{2}{Astronomy Program, SEES, Seoul National University, Seoul,
 151-742, Korea}
\altaffiltext{3}{Departamento de Fisica, Universidad de Conception,
Casilla 160-C, Conception, Chile}

\begin{abstract}

We present a systematic study of the low mass X-ray binary (LMXB)
populations of 6 elliptical galaxies, aimed at investigating the
detected LMXB $-$ globular cluster (GC) connection.  We utilize
{\chan} archival data to identify 665 X-ray point sources and {\hst}
archival data supplemented by ground observations to identify 6173
GCs. Applying rigorous X-ray and optical photometry and conservative
matching criteria, we associate 209 LMXBs with red GC (RGC) and 76
LMXBs with blue GCs (BGC), while we find no optical GC counterpart for
258 LMXBs.  This is the largest GC$-$LMXB sample studied so far.

We confirm previous reports suggesting that the fraction of GCs
associated with LMXBs is $\sim3$ times larger in RGCs than in BGCs,
indicating that metallicity is a primary factor in the GC$-$LMXB
formation.  While as already known, the brighter (and bigger) GCs have
a higher probability to host LMXBs, we find that this optical
luminosity (or mass) dependency is stronger in RGCs than in BGCs.  We
also find that GCs located near the galaxy center have a higher
probability to harbor LMXBs compared to those in the outskirts.  The
radial distributions of GC$-$LMXBs (for both RGC and BGC) are steeper
than those of the whole optical GC sample, but consistent with those
of the optical halo light, suggesting that there must be another
parameter (in addition to metallicity) governing LMXB formation in
GCs. This second parameter must depend on the galacto-centric
distance. One possibility is a galacto-centric distance dependent
encounter rate.

We find no statistically significant difference in the X-ray
properties (shape of X-ray luminosity function, $L_X/L_V$
distribution, X-ray spectra) among RGC$-$LMXBs, BGC$-$LMXBs and
field$-$LMXBs. The similarity of the X-ray spectra of BGC$-$LMXBs and
RGC$-$LMXBs is inconsistent with the irradiation-induced stellar wind
model prediction of more absorbed X-ray spectra in BGC$-$LMXBs than in
RGC$-$LMXBs. The similarity of the X-ray luminosity functions (XLFs)
of GC$-$LMXBs and field$-$LMXBs indicates that there is no significant
difference in the fraction of BH binaries present in these two
populations, in contrast to Galactic LMXBs where BH binaries are not
found in GCs. The similar X-ray properties as well as the similar
radial distributions of GC$-$LMXBs and field$-$LMXBs cannot constrain
the hypothesis that all LMXBs were formed in GCs.

\end{abstract}

\keywords{X-ray:Low Mass X-ray Binaries --- galaxies: elliptical}

\section{Introduction}

{\chan} observations of nearby early-type galaxies have shown that a
large fraction (20-70\%) of LMXBs are associated with GCs (e.g.,
Sarazin, Irwin, \& Bregman 2001; Angelini, Loewenstein, \& Mushotzky
2001; Kundu, Maccarone, \& Zepf 2002; Minniti et al. 2004; and Jordan
et al. 2004; see also an archival study by Sarazin et al. 2003 and a
recent review by Verbunt \& Lewin 2005).  Moreover, a study of LMXBs
near the center of NGC 4472 showed that the metal-rich red GCs (RGCs)
are more likely by a factor of $\sim3$ to harbor LMXBs than blue GCs
(BGCs) (Kundu, Maccarone, \& Zepf 2002). This preferred association of
LMXBs with metal-rich GCs (but with some variations among galaxies)
was later confirmed by Sarazin et al. (2003) and Jordan et al (2004),
indicating that metal abundance plays a key role in LMXB formation; a
similar trend was also known in the Milky Way (Grindlay 1993;
Bellazzini et al. 1995). Several binary formation/evolution scenarios
have been discussed to explain these results (e.g., Grindlay 1993;
Bellazzini et al. 1995; Maccarone, Kundu, \& Zepf 2004; Ivanova 2005);
however, it is still unclear why metallicity plays such an important
role in LMXB formation and what causes the observed galaxy to galaxy
variations in the fraction of LMXBs associated with RGCs and BGCs.

The LMXB $-$ GC association is particularly intriguing, because the
high stellar density near the center of GCs may trigger the formation
of binaries effectively by either three-body process or tidal capture.
These binaries would then evolve into LMXBs, as first suggested by
\citet{gh85} for the Milky Way. Given the larger density of GCs in
elliptical galaxies \citep{harris91}, this mechanism has been recently
reproposed \citep{sib01,white02}. The observational evidence is,
however, ambiguous. The majority of LMXBs are not directly connected
to GCs. The field$-$LMXBs might have evolved from field binary stars
or might have been formed in GCs and then dispersed in the field;
either SN kick or dynamical interaction within GCs have been
considered as ways to extract LMXBs from their parent GC
\citep{white02}. Recently, \citet{juett05} and \citet{irwin05} analyzed
the relation between the fraction of GC$-$LMXBs (or their co-added X-ray
luminosity of LMXBs) and the GC specific frequency ($S_N$) and
concluded that exclusive GC formation for LMXBs may not be supported
by the data.  However, all these observational studies are based on
relatively small GC$-$LMXB samples, resulting from the small field of
view of the {\hst} observations used to identify GCs.

Here, we utilize both {\hst} and ground-based observations in our study
of 6 elliptical galaxies observed with {\chan}. \citet{lk00} showed
that the {\hstwfpc2} data supplemented by wide-field ground based data
are very effective to investigate GC systems of elliptical galaxies,
since the crowding problem in ground-based observations is not severe
for the outer region of the galaxy. Combination of the {\hst} and
ground-based observations enables us not only to expand the available
sample, but also to extend the study of LMXBs to the outer part of the
galaxy, which has not been studied so far. It is also important to
compare LMXBs and GCs in the same galaxy scale and extract their
properties (e.g., $L_X$(LMXB), $f$(GC-LMXB) and $S_N$) from the same
region.

This paper is structured as follows: In \S 2 we describe the
sample galaxies and the data analysis method for the extraction of
both LMXB and GC samples.  We also explain in detail how GC candidates
are selected and discuss the degree of contamination in the optical GC
sample. The match between X-ray and optical source lists and the
accuracy of astrometry are described in \S 3. We report our main
results in \S 4.  We discuss the comparisons between different
LMXB populations and their implications in \S 5 and summarize our
main results in the last section.

\section{Sample Selection and Basic Data Reduction Techniques}

\subsection{The Sample}

We selected relatively nearby (d $\le 30Mpc$) elliptical galaxies with
archival {\chan} ACIS
data centered on the \s3 chip \citep{weisskopf00} to study the
characteristics of X-ray point sources. Moreover, since our goal is to
investigate not only the X-ray properties of the LMXBs but also the
LMXB $-$ GC connection, we required that the sample galaxies had been
observed both in optical and X-ray wavelengths.  In addition, to
optimize the optical coverage of our sample, we only considered
galaxies where both ground-based wide-field observations and central
{\hstwfpc2} pointings are available. Our final sample contains six
galaxies: NGC 1399, NGC 4374, NGC 4472, NGC 4486, NGC 4636 and NGC
4649.  Five of these six galaxies are in the Virgo cluster.  We list
the target names along with the positions and the basic photometric
properties in Table \ref{tab-target}.

In all cases, the ACIS$-$S3 CCD was primarily used to take advantage
of the higher sensitivity at the low energies (kT $<$ 4 keV) of the
back-illuminated S3 chip \citep{pog04}. To have homogeneous X-ray
data, we restricted the analysis to only data from the \s3 chip even
though there are still some point sources detected in the other CCDs.
The obs-id and the exposure times of the {\chan} observations used are
listed in Table \ref{tab-xobs}.  The effective exposure times range
from $22$ksec(NGC 4649) to $100$ksec(NGC 4486).

Fig. \ref{fig-fov} shows the ground based optical images of the
6 galaxies. The big rectangles represent the boundary of the {\chan} \s3
chip; the bat-shaped regions give the field of view of the
{\hstwfpc2} observations, and the ellipses show the optical extent of the
galaxy based on the standard diameter ($D_{25}$) and ellipticity ($\epsilon$)
from the RC3 catalog \citep{deV91}.
The point sources detected in the \s3 chip
are marked with small circles, where
the size of the circle is proportional to the size of the {\chan}
{\psf} ($95\%$ of the encircled
energy fraction).

Ground-based optical observations for the 6 elliptical galaxies were
carried out with the 4$m$ telescopes at Kitt Peak National Observatory
(KPNO) and Cerro Tololo Inter$-$American Observatory (CTIO). We used
the Washington $C$ and standard Johnson $R$ filters. The Washington
filter system (e.g., $C$ \& $T_1$ filters) is known to be effective in
discriminating the GCs from other contaminating sources, such as faint
background galaxies and foreground stars \citep{glk96}. We note that
the $R$ magnitude is similar to the $T_1$ magnitude \citep{klg00,dirsch03},
while the $R$
filter has $\sim3$ times the sensitivity of the $T_1$ filter
\citep{geisler96}. We list the journal of the optical ground-based
observations of the target galaxies in Table \ref{tab-gobs}.

While deep ground-based optical observations provide a large enough
field of view to cover the whole \s3 chip, they suffer from
saturation and crowding especially in the central part of a galaxy.
For these central regions, we use archival
{\hstwfpc2} data.  We use $V$ and $I$ band images, since these bands
have been well calibrated to identify GC candidates in external galaxies
\citep[e.g.,][]{kissler00}.  We take the $F547M$, $F555W$ and $F606W$
filters for the $V$ band and the $F814W$ filter for the $I$ band.  The
list of archival {\hstwfpc2} observations for the 6 galaxies is given in
Table \ref{tab-hobs} with exposure times and the field observed
shown in Fig. \ref{fig-fov}.  The filter name for $V$ band is
specified when either $F547M$ or $F606W$ are used instead of $F555W$.
We only list the total exposure times of multiple exposures.  The mean
exposure times for $V$ and $I$ bands of {\hstwfpc2} observation are
$1950$ sec and $1970$ sec, respectively.

\subsection{X-ray and Optical Photometry}

Following the X-ray data reduction procedure in the {\chan}
Multiwavelength Project (ChaMP) \citep{kimetal04}, we clean 
background flare events during the exposure, then detect point
sources with the CIAO task {\it wavdetect}.  The X-ray photometric
properties (count rate, hardness ratio, etc.) of point sources detected
in the $B$ band are computed by adding events in a circular aperture
corresponding to $95\%$ of the encircled energy of the
{\chan} point spread function\footnote{http://cxc.harvard.edu/cal/Hrma/psf}.

\begin{figure}
\epsfig{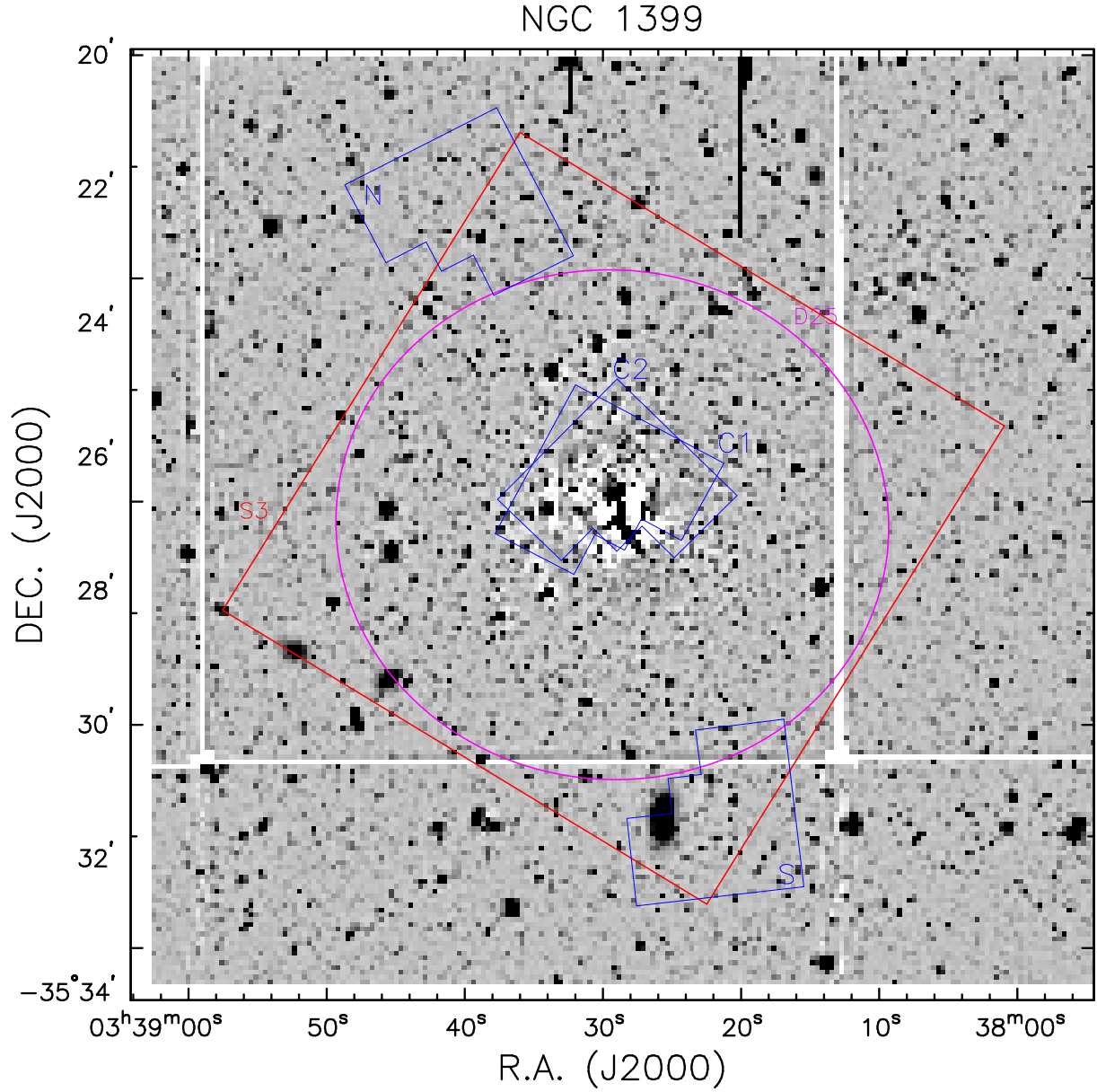} 
\vspace{-3mm}
\caption{Ground-based $C$ filter image of NGC 1399. Galaxy halo light is
removed using ellipse fitting and median filtering method
(see text for more explanation).}
\label{fig-fov2}
\end{figure}

We follow several approaches to obtain optical photometry from the
ground-based observations. For NGC 4472, we use the $CT_1$ photometry
of \citet{glk96} and \citet{lkg98}. Since the $R$ magnitude is very close
to the $T_1$ magnitude \citep{geisler96}, we simply consider the $T_1$
magnitude for point sources in NGC 4472 as their $R$ magnitude. For
NGC 1399, we use the wide field photometry of
\citet{dirsch03}.  For the remaining four galaxies, we use the
photometric data of \citet{lee05}.  To decrease the effect of highly
varying galaxy halo light on the detection and brightness of point
sources, the halo light is modelled with IRAF/STSDAS median smoothing
at the outer radii and ellipse fitting tasks in the inner regions of
the galaxies. Except for the very central region where saturation
truncated source detection in all ground-based observations, this
modelling is very successful in reducing the effect of the galaxy
light. As an example, Fig. \ref{fig-fov2} shows the residual image of
NGC 1399 after subtracting the model diffuse emission.

We use the \citet{lk00} {\hstwfpc2} photometry of NGC 4472; these
authors used simple circular aperture along with appropriate aperture
correction to obtain the photometry of point sources.  For the
remaining five galaxies, we perform our own
{\hstwfpc2} photometry in this study. Images obtained with the same
filter are combined if there are multiple images, to remove cosmic ray
hits. Rejection of cosmic rays is important, especially for the
{\hstwfpc2} observation. Because the {\hst} point spread function is
comparable to the size of cosmic ray hits, a cosmic ray is more likely
misidentified as a valid point source in the {\hst} observations than in
ground-based observations.  To remove the galaxy diffuse emission, we
apply the same modeling technique as used for the ground-based
observations.  For the final photometry, we utilize the digital
photometry software HSTPHOT \citep{dolphin00,kimetal02}.



\begin{figure}
\epsfig{figure=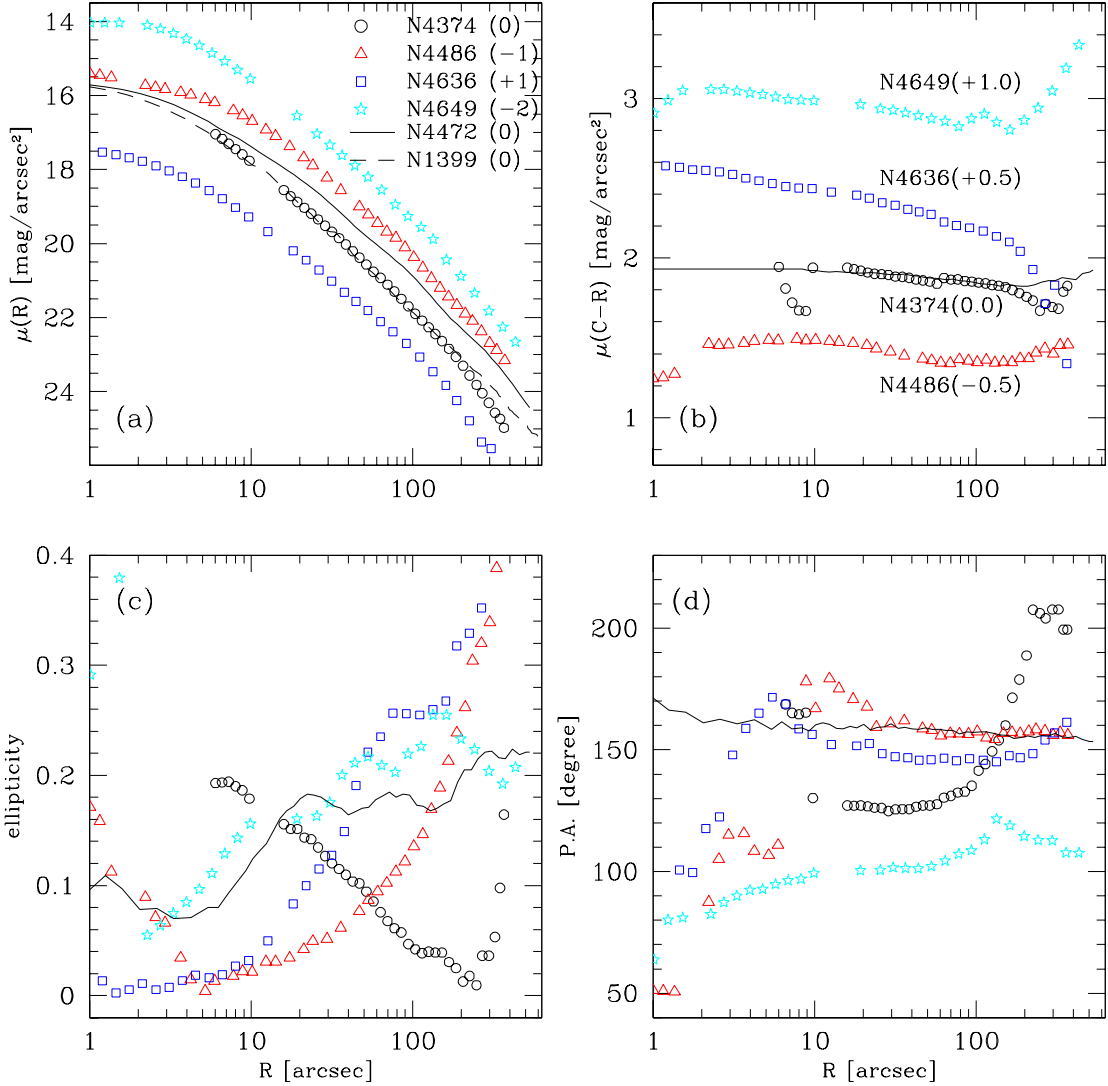, height=0.975\textwidth, width=0.975\textwidth}
\vspace{-3mm}
\caption{Surface photometry of 4 elliptical galaxies: radial profiles of
(a) $R-$band surface brightness
(b) $(C-R)$ surface color
(c) ellipticity in the $R-$band image and
(d) position angle in the $R-$band image.
NGC 4472 \citep{klg00} and NGC 1399 \citep{dirsch03} are also shown
for comparison with solid lines and
dashed lines, respectively.}
\label{fig-sphot}
\end{figure}

For the four galaxies NGC 4374, NGC 4486, NGC 4636 and NGC 4649, we
obtain surface photometry of images from ground-based observations
using the {\it ellipse} task of
IRAF/STSDAS. Iteratively fitting an ellipse to isodensity contours
\citep{jed87}, {\it ellipse} provides radial profiles of brightness,
color, ellipticity and position angle. The ellipse fitting results are
shown in Fig.  \ref{fig-sphot}. For comparison, we also plot the NGC
4472 (solid lines; Kim, Lee, \& Geisler 2000) and NGC 1399 (dashed lines;
Dirsch et al. 2003).  Based on the $R-$band surface photometry, we list in
Table \ref{tab-sphot} the basic structure parameters, including
effective radius ($R_{eff}$) and standard radius ($R_{25}$),
a circular radius of an ellipse where the surface
brightness in $B-$band is $\mu_B = 25$.  Also
listed are ellipticities, position angles, and colors at these two
radii.  We measure the mean color and magnitude and color gradient in
the region of $R_{eff} \le R \le R_{25}$ (the last three rows in Table
\ref{tab-sphot}). We note that the mean optical properties are similar
in different galaxies, with the possible exception of NGC 4636 which
has a slightly bluer color in the outer radii.

\subsection{Optical Globular Cluster Candidates}

\begin{figure}
\epsfig{figure=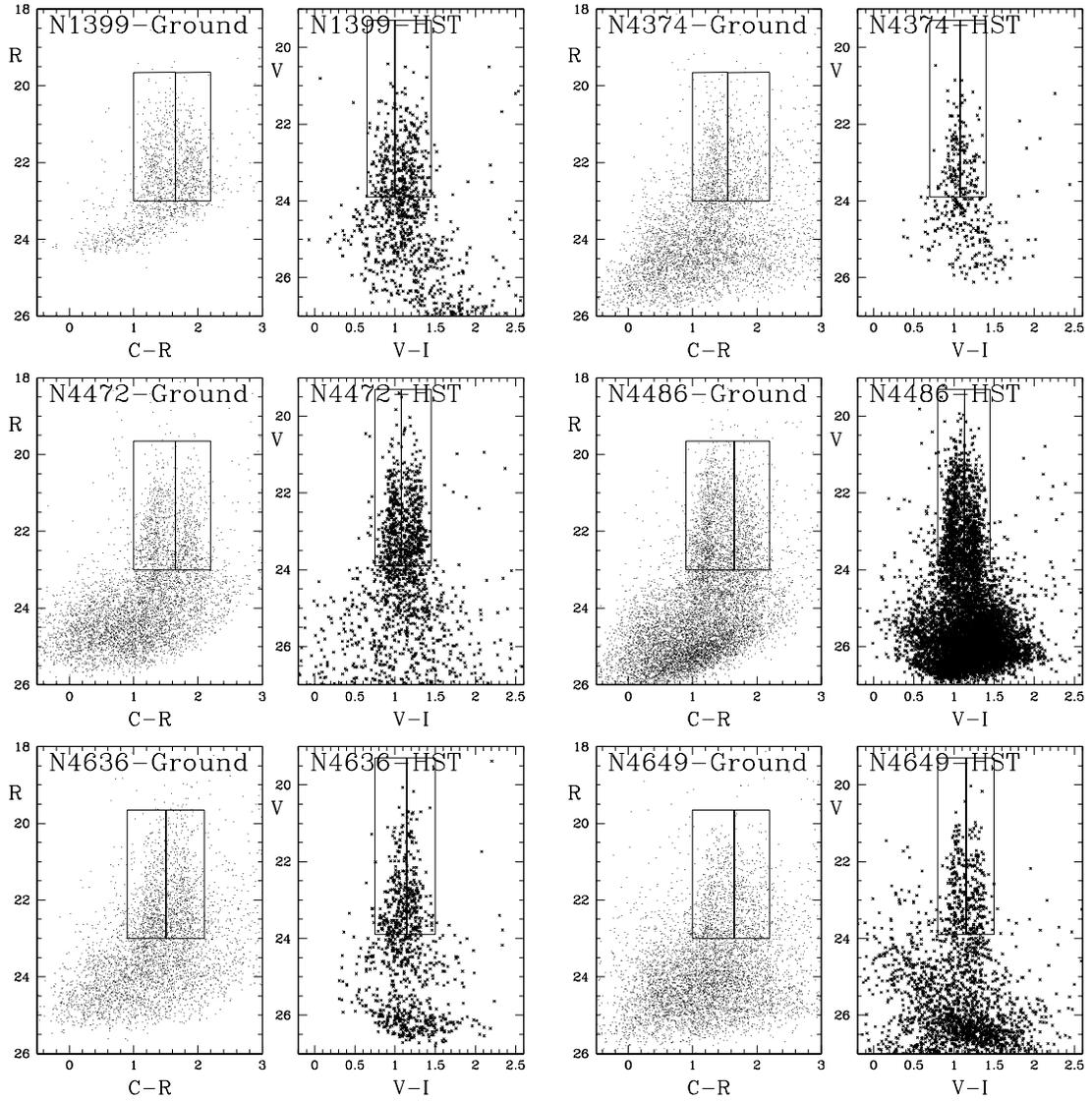, height=0.975\textwidth, width=0.975\textwidth}
\vspace{-3mm}
\caption{CMDs of point sources detected in ground
and {\hstwfpc2} observations. The selection boundaries of GC candidates
are shown with rectangles (see text).}
\label{fig-cmd}
\end{figure}

Multi-color observations are frequently used for selecting GC
candidates, since GCs show a typical distribution in a color magnitude
diagram(CMD), especially in the $(V-I)-V$ and $(C-R)-R$ domains
\citep{glk96,lk00}.  Fig. \ref{fig-cmd} shows the
distribution of optical point sources in the $(V-I)-V$ CMD for
{\hstwfpc2} observations and in the $(C-R)-R$ CMD for the ground-based
observations of the six elliptical galaxies. The boxes represent the
selection criteria of GC candidates adopted in the present
study. Since most of the sample galaxies are in Virgo, we used the
same magnitude boundaries defined by \citet{glk96} for ground-based
observations ($19.65 < R < 23.5$) and by \citet{lk00} for the
{\hstwfpc2} observations($V < 23.9$).  We determined the ($V-I$) and
($C-R$) color boundaries by inspecting the color distribution of point
sources at different magnitude limits and listed them in Table
\ref{tab-gcbox}.  Since point sources detected in {\hstwfpc2}
observation are more reliable than those from ground-based
observations, we preferentially use {\hstwfpc2} data whenever
possible. $CR$ photometries are transformed to $VI$ photometries by
using point sources detected in both {\hstwfpc2} and ground-based
observations.  The total number of GC candidates in the six galaxies
is 6173 in a radial region of $20'' < R < R_{25}$.  M87 (NGC 4486) is
found to have the most populous GC system ($N_{GC} = 1906$)
while NGC 4374 has only 523 GCs.  In Table \ref{tab-psrc1}, we
summarize the number of GCs found in ground and {\hst} observations.


Even with the above selection criteria, a significant fraction of
interlopers remain in the selected GC candidates, since there are some
background galaxies which have magnitudes and colors located in the
selection boxes of the CMDs of Fig. \ref{fig-cmd}.  The sources
detected in both {\hstwfpc2} and ground-based observations enable us to
estimate the number of these contaminants. We assume that (1) there
are no contaminants in the sample of point sources detected in the
{\hstwfpc2} observation and (2) that all the GCs detected in
ground-based observation are also detected in the {\hstwfpc2}
observation. The first assumption is quite plausible, since the high
spatial resolution of {\hst} enables us to reliably distinguish a point
source from a background galaxy. The second assumption requires some
scrutiny, since it depends not only on the observational status
(e.g. exposure time, filter system), but also on the distribution of
sources within the galaxy. Based on artificial source experiments,
\citet{lk00} showed that in the inner region of NGC 4472, the
incompleteness due to the highly varying galaxy light and source
crowding is negligible only when bright ($V \le 24$ mag) sources
outside the crowded center ($r \ge 10$ arcsec) are used. Because our
sample covers a wide range of exposure times, we apply
conservative criteria to validate the second assumption: $r > 20$
arcsec and $V \le 23.9$ mag.

\begin{figure}
\epsfig{figure=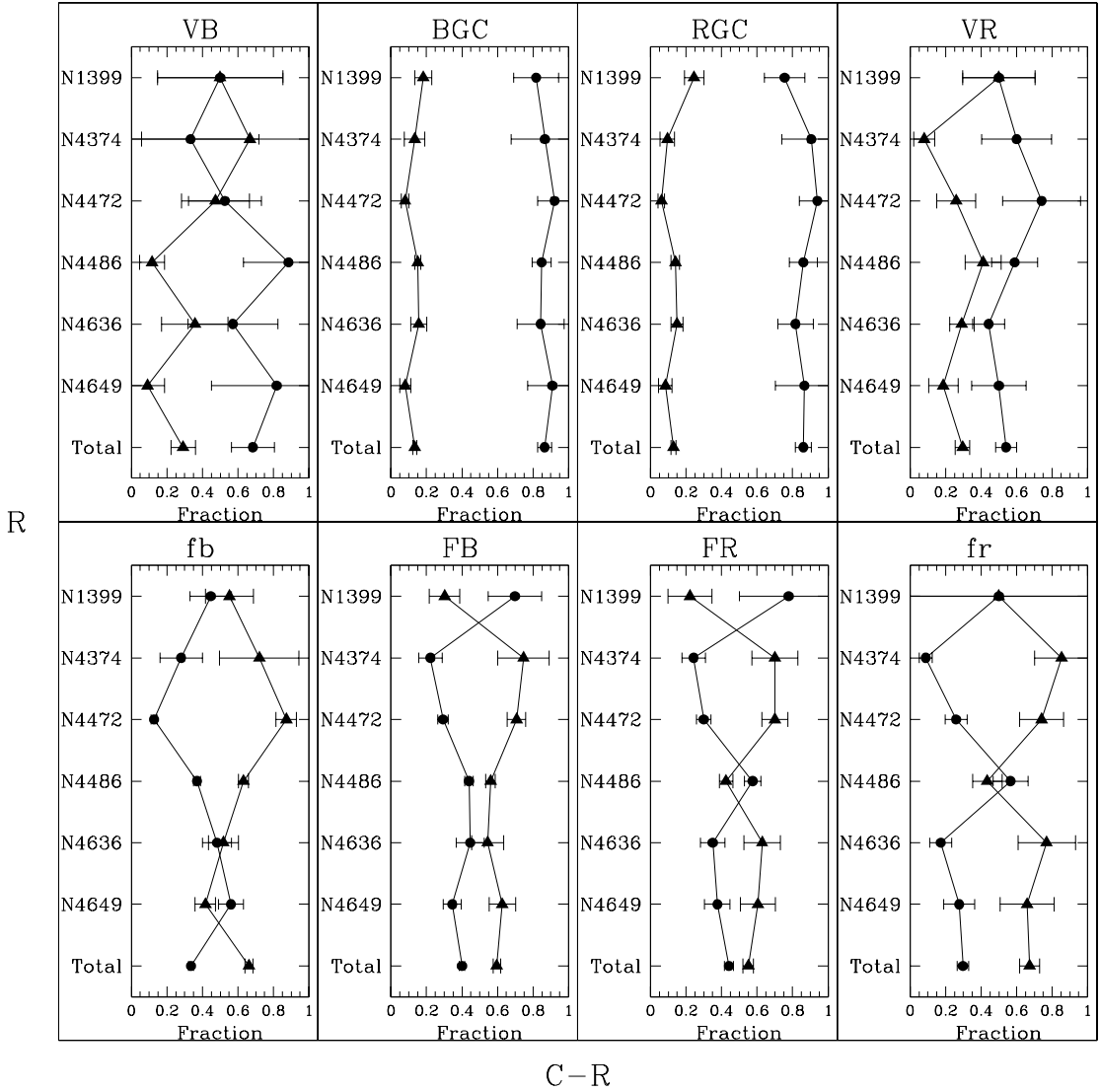, height=0.975\textwidth, width=0.975\textwidth}
\vspace{-3mm}
\caption{Ratios of point sources detected in ground-based observation
and {\hstwfpc2} observations.
The panels designated as BGC and RGC represent blue and red GCs,
respectively. The CMD region VB (very blue; upper-left panel)
represents sources with a bluer color than the color of blue GCs but
with the same magnitude range. Similarly, VR (very red; upper-right
panel) represents redder sources. FB (faint blue) and FR (faint red)
in the lower panel indicate points sources fainter than blue and red
GCs, but with the same colors.
We plot the ratio $N_2/N_t$ with filled circles connected with solid line,
and the fraction of contaminants ($N_1/N_t$) with filled triangles
(See text for the definition of $N_1, N_2$ and $N_t$).}
\label{fig-cont}
\end{figure}

We show the number ratio of the detected point sources located in a
given CMD-region in Fig. \ref{fig-cont}.  This ratio is defined as
$N_t = N_1 + N_2$, where $N_1$ is the number of point sources detected
only in ground-based observations, and $N_2$ the number of point
sources detected in both ground and {\hst} observations.  Although the
degree of contamination varies slightly from one galaxy to another,
the global amount of contaminants for the sample of GC candidates in
ground-based observations is $\sim14\%$ for CMD regions BGC \& RGC.
The contamination increases dramatically when we consider the fainter
samples (CMD regions FB \& FR in Fig. \ref{fig-cont}) or sources with extreme
colors (CMD regions VB \& VR in Fig. \ref{fig-cont}).

%
%
%
%
%

Applying this contamination fraction to the GC candidates selected
from ground observations alone, we estimate the number of contaminants
in BGC and RGC candidates of ground-based observation at different
galacto-centric radii and list the numbers in the last two columns of
Table \ref{tab-psrc1}.  The total number of background galaxies which
might be included in the current optical globular cluster sample
($20'' \le R \le R_{25}$) is 9\%. The contamination increases with
galacto-centric radii due to the lack of {\hstwfpc2} observations in the
outer part of a galaxy. The fraction of contaminants for the inner
regions (outside the central $20''$) is less than $5\%$, while
this value increase to $\sim12\%$ for the region outside the \D25
ellipse.

As an independent test to address the contamination in the sample of
ground based observations, we utilize the source catalogs from Hubble
Deep Fields north and south \citep{hdfn,hdfs}. Since the typical
seeing size of ground based observations is $\sim1''$, we only consider
extended sources with a size smaller than $1''$, which might be
misclassified as a point source in ground based observations. Counting
the number of HDF sources located in the same CMD region with our
optical GCs, we find that the misclassified extended galaxies could be
$\sim13\%$ of our ground-based optical GC sample, corresponding to
$\sim8.4\%$ of our GC sample.  We conclude that the effect of
contamination in our GC sample is small ($< 10\%$).

\section{Matching X-ray and Optical Positions}

\subsection{Astrometric Uncertainty}

To accurately determine the X-ray source position, we first corrected
for the aspect offset introduced by the earlier inaccurate calibration
data (see CIAO science thread\footnote{
http://asc.harvard.edu/ciao/threads/arcsec\_correction/}).
After running {\it wavdetect}, we also applied a position
refinement algorithm \citep{kimetal04}, which was later incorporated
into
CIAO v3.0 {\it wavdetect}\footnote{http://cxc.harvard.edu/ciao}.
The X-ray source positional error is then calculated by
the prescription given by \citet{kimetal05}, who provide a set of empirical
equations as a function of source count and off-axis angle, based on
extensive simulations. The positional uncertainty of a typical X-ray
source with 30 net counts at off-axis angle $4$ arcmin is
$0.83$ arcsec at the $95\%$ confidence level.

The optical source position in pixel coordinates is determined by
{\it daophot} \citep{stetson87} for ground-based observations and HSTPHOT
\citep{dolphin00}
for {\hstwfpc2} observations, respectively. The pixel coordinate is
then transformed to the world coordinate by cross-correlating the
USNOB1 catalog \citep{monet03}. A typical error of this transformation is
$\sim0.3$ arcsec. However, we cannot use the USNOB1 catalog for the
{\hstwfpc2} data because there are only a small number of USNOB1 sources
inside the {\hstwfpc2} field of view, particularly when it is aiming at the
center of a galaxy which is often saturated and also highly non-uniform.
Instead, we use the transformed ground-based data as a template to
obtain the world coordinate for {\hstwfpc2} sources. The transformation
error between the ground-based data and {\hstwfpc2} data is dominated by
the former error ($\sim0.3 ''$) due to the negligible positional
error of the {\hstwfpc2} data.

\subsection{Match and Random Match}

Using the X-ray positional errors at the 95\% confidence level
\citep{kimetal05}, we first select optical sources inside the {\chan}
error radius. To minimize false matches, we further limit the search
radius to $1.2 ''$ and visually inspect the optical image for
validation. Most ($> 91\%$) X-ray sources have error radii smaller
than $1.2''$.  When there are multiple optical sources inside the
search radius, we select the nearest optical source. Since this
happens only 11 times (or for a few \% of the sources),
the expected number of false
matches is negligible.  For $85\%$ of the matches, the positional
offset between X-ray and optical sources ($d_{XO}$) is less than the
typical positional error of X-ray sources ($0.83''$). The median
positional offset is $d_{XO} = 0.5 ''$.

To identify field LMXBs (i.e., X-ray sources without an optical
counterpart), we first select the X-ray sources without optical
counterpart inside $d_{XO} = 1.2''$.  We then visually inspect the
optical images at the position of X-ray sources to look for optical
counterparts just outside the error circle and undetected sources due
to chip defects and/or saturation effect of nearby bright sources.  In
our sample of 665 X-ray point sources inside the \D25 ellipse (but
excluding the central 20 arcsec region), we identify 285 ($43\%$)
GC$-$LMXB and 258 ($39\%$) field$-$LMXB. We list the match statistics
for each galaxy in Table \ref{tab-psrc2}.  Since our match criteria
are rather conservative, we are left with 122 X-ray sources for which
we cannot establish whether they match with GCs or not.  Some of them
match with non-GC optical sources. We do not use these X-ray sources
in the following analysis to minimize the uncertainty of X-ray source
characteristics.

To assess the probability of false matches, we applied two independent
methods. First, to determine the probability of finding an optical
source by chance, we shifted the X-ray source position by $5''$,
corresponding to $\sim6$ times the typical 95\% position error, and
then we tried to match them with optical sources as explained above.
Based on 100 simulations for the whole \s3 field of view, we find
$\sim17 \pm 1 (\sim2.5\%)$ matches, which is considerably smaller
than the 43\% match occurrence obtained for the observed X-ray source
positions. Secondly, we redistributed the optical sources randomly,
but following the observed GC radial profile, and kept the number of
sources the same as that of the observed sources. The number of random
matches with optical GCs is $N=34 \pm 4$ (or $\sim5\%$), and
decreases (by $\sim30\%$) when X-ray sources in $r < 20$ arcsec are
excluded. Adopting the radial profile of the halo light instead of the
GC profile affects very little the number of random matches ($N=41$).
Therefore, we conclude that the chance probability of a false match is
small compared to the number of GC$-$LMXB associations.

We summarize the match statistics for the point sources in this radial
range in Table \ref{tab-psrc3}.  The mean probability for a GC to
harbor an LMXB, defined by $f_{GC}$ = N(GC$-$LMXB) / N(GC), is $\sim$
5.2\% in our sample of 6 galaxies.  This is slightly larger than that
of \citet{kmz02} for the central region of NGC 4472, mainly due to the
high fraction (9.8\%) in NGC 1399. The mean probability excluding NGC
1399 is $4.4\%$.  While the fraction of BGCs with an LMXB ($f_{BGC}$)
is relatively constant ($\sim$ 2\%) with the exception of
NGC 1399 (5.8\%), that of
RGCs ($f_{RGC}$) widely varies from one galaxy to another (2.7\% --
13\%), resulting in $f_{RGC}/f_{BGC}$ varying from 1.4 to 4.6.  The
average $f_{RGC}/f_{BGC}$ is 2.7, indicating that RGCs on average have
a higher probability to harbor an LMXB by a factor of 2.7 than BGCs.
Although the mean value is consistent with those previously reported
\citep[e.g.,][]{kmz02,sarazin03}, we note that galaxy to
galaxy variations are not negligible.

\begin{figure}
\epsfig{figure=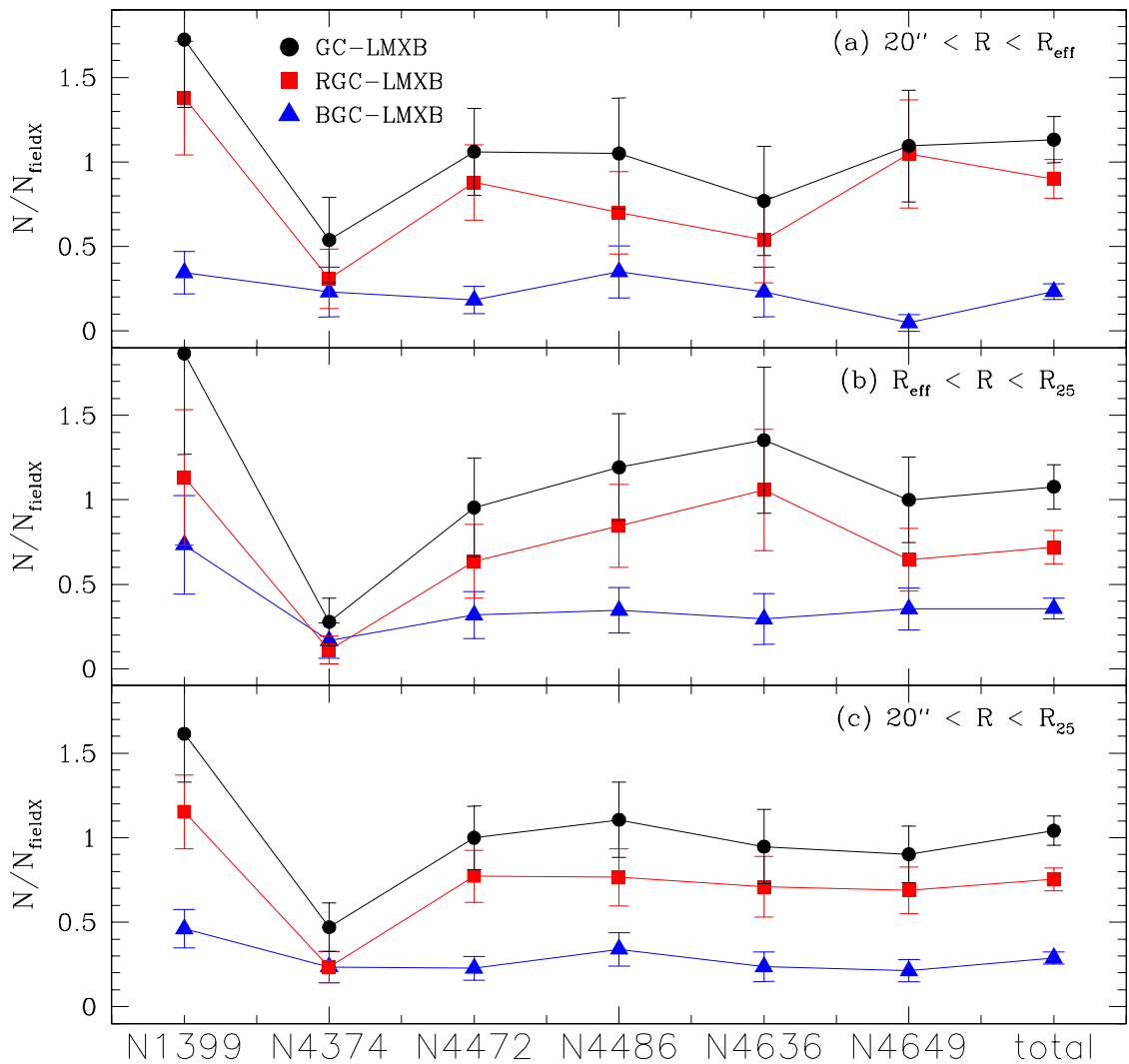, height=0.975\textwidth, width=0.975\textwidth}
\vspace{-3mm}
\caption{Ratio of N(GC$-$LMXB) to N(field$-$LMXB) for six galaxies.
The symbols of filled squares, filled circles,
and star marks represent the ratios for BGC$-$LMXB, RGC$-$LMXB, and
GC$-$LMXB, respectively.}
\label{fig-n2nx}
\end{figure}

The total number ratio of N(GC$-$LMXB) to N(field$-$LMXB) is close to
$1$.  We show the run of this ratio as a function of individual galaxy
in Fig. \ref{fig-n2nx}. The most extreme ratios are found in NGC
4374 and NGC 1399.  The low ratio of NGC 4374 ($\sim0.4$) appears to
be mainly due to the small number of the RGC$-$LMXB population, while
the higher ratio for NGC 1399 ($\sim1.8$) is due to the small number
of field$-$LMXBs in this galaxy. Again we note that galaxy to galaxy
variations are not negligible (see also \S 5).



\section{Comparison between BGC$-$LMXB, RGC$-$LMXB and Field$-$LMXB}

\subsection{GC Luminosity Distribution}

\begin{figure}
\epsfig{figure=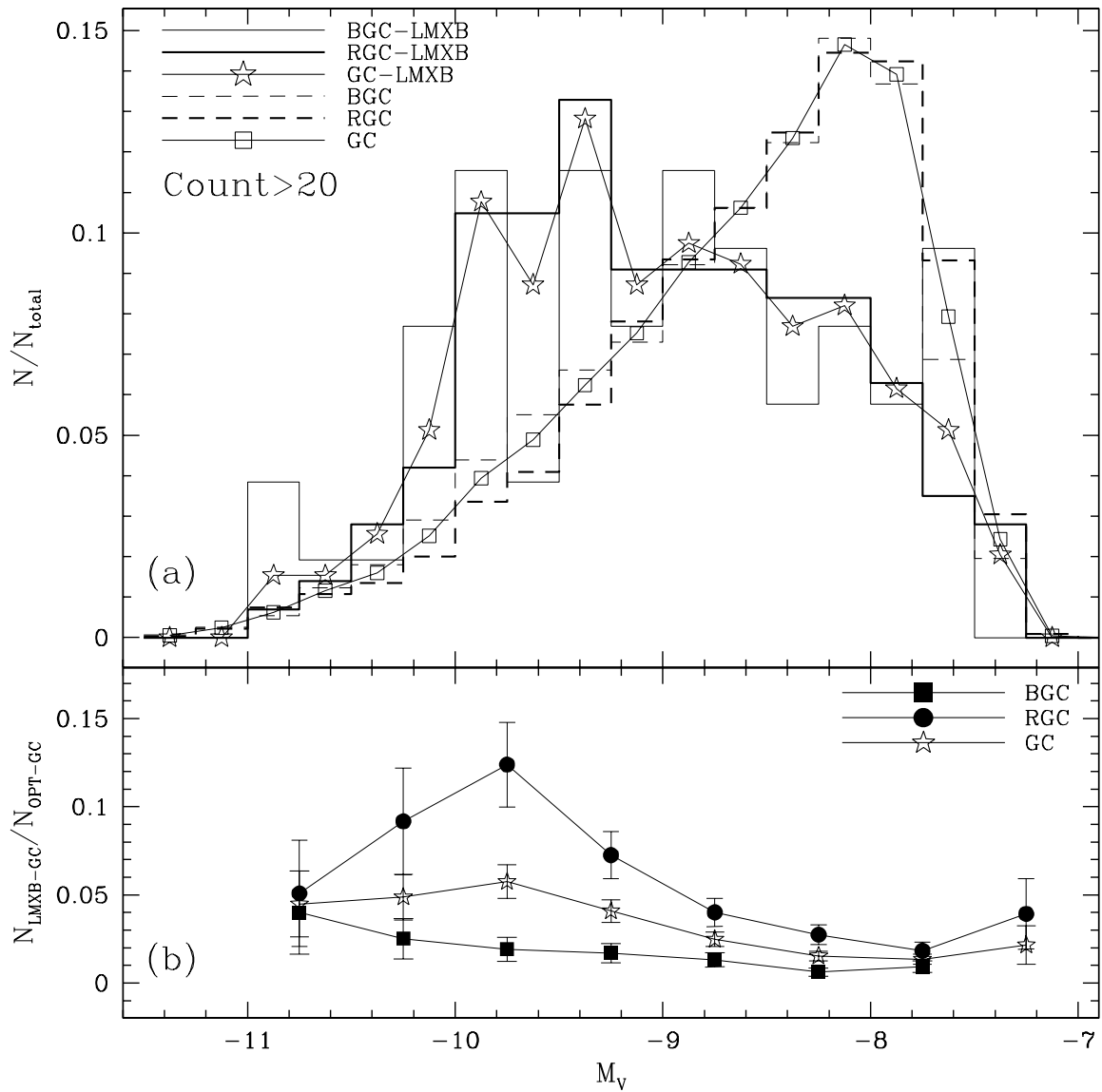, height=0.975\textwidth, width=0.975\textwidth}
\vspace{-3mm}
\caption{(a) Optical luminosity function of GCs associated with LMXBs
and the entire GC sample, (b) Ratio of luminosity functions of
GCs with LMXBs and entire globular clusters.}
\label{fig-olum}
\end{figure}

We show the optical luminosity ($M_V$) distributions of GCs (the whole
sample and red/blue GCs separately) in Fig. \ref{fig-olum}.  We also
plot the luminosity distribution of GCs with LMXBs,
for which we use only the LMXBs with net count $>20$
to minimize the effect of incompleteness, i.e., missing faint X-ray
sources. In this plot, we use sources outside of central $20''$ radius.


The amplitude of the observed optical luminosity function of GCs
increases as the luminosity
decreases and peaks at $M_V \approx -8$. Beyond $M_V \approx -8$ the
observed luminosity function drops quickly due to incomplete detection.
But note that the true peak of the luminosity function is at
$M_V \approx -7.5$ \citep{richtler03}.
The luminosity function of GCs with LMXBs is different from that
of the whole GC sample: the observed peak luminosity is
significantly brighter than that of the whole GC sample by $\sim1.5$
mag (or $\sim4$ times brighter), indicating that brighter (bigger) GCs
preferentially harbor LMXBs.  This is consistent with previous
results \citep[e.g.,][]{sarazin03}. Applying the Kolmogorov-Smirnov (KS)
test for the sub-samples used in Fig. \ref{fig-olum}, we find that the
probability that GCs with LMXBs are drawn from the same parent
population as the whole GC sample is negligible ($< 1\%$).

The ratio of the luminosity distribution of GCs (both total, blue and
red) with LMXBs and the whole GC samples is shown in
Fig. \ref{fig-olum} (b). Again, there is a clear indication that
brighter (bigger) GCs preferentially harbor LMXBs.  Interestingly,
this trend of the optical luminosity dependency is much stronger in
RGCs than in BGCs. For example, the fraction of RGCs with LMXB at $M_V
= -9.8$ is $\sim5.5$ times higher than that at $M_V = -8.0$, while the
fraction of BGCs with LMXBs only changes by a factor of $\sim2$ (see
\S 5 for more discussion).  The KS-test weakly rejects (at the $90\%$
confidence level) the hypothesis that BGCs and RGCs which harbor
LMXBs belong to the same population.

\subsection{Radial Distributions of GCs}

Our large sample of sources and the larger optical field of view of
the ground-based data allow, for the first time, the studies of
the radial variations of GC/LMXB properties, including the optical
luminosity of GCs hosting LMXBs, and X-ray luminosities and colors of
GC$-$LMXBs. Previous work, because of the small {\hst} field of view,
could not extend the study of GC $-$ LMXB associations farther than $r
> {\approx 3'}$, where $r$ is the galacto-centric radius. 

\begin{figure}
\epsfig{figure=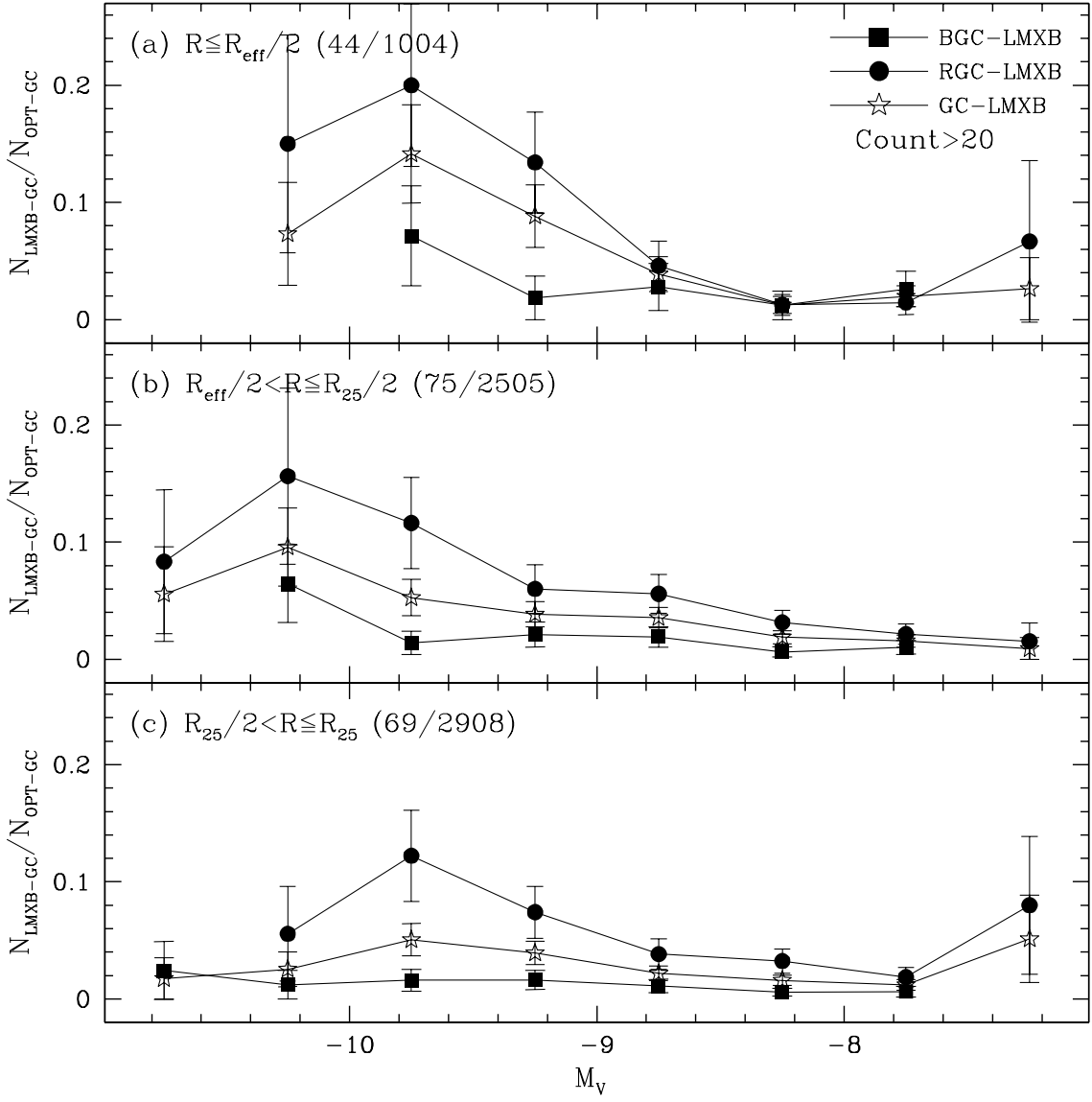, height=0.975\textwidth, width=0.975\textwidth}
\vspace{-3mm}
\caption{Ratio of luminosity functions of GCs with LMXBs and the entire GC for
(a) the central region (b) the intermediate region, and (c) the outer regions.
We use the same symbols as in Fig. \ref{fig-olum}(b).
The numbers in parenthesis are the number of
GCs with LMXBs and the total number of GCs in each radial region.}
\label{fig-rolum}
\end{figure}

Fig. \ref{fig-rolum} shows the number ratio of GC$-$LMXBs/GCs for different radial
regions: (a) $R \le R_{eff}/2$, (b) $R_{eff}/2 < R \le R_{25}/2$ and
(c) $R_{25}/2 < R \le R_{25}$, where $R_{eff}$ and $R_{25}$ are from
Table 5.
We consider
only LMXBs with net counts $> 20$. 
The fact (\S 3.2) that RGCs more preferentially harbor
LMXBs than BGCs remains valid in all three radial bins. Also valid is
the fact (discussed in \S 4.1) that the brighter GCs more
preferentially harbor LMXBs. However, it is interesting to note that
both RGCs and BGCs located near the galaxy centers have a higher
probability to harbor LMXBs compared to the GCs at outskirts. The
enhanced probability is most significant for the bright ($M_V < -9$)
RGCs, reaching $\sim20\%$ at the peak in the innermost radial bin. In
the outer regions, this probability goes down to $\sim10\%$.  For
faint GCs, the radial difference is less clear since the number of GCs
with LMXBs is small. This radial dependency, which is reported here
for the first time, seems to suggest an important mechanism for LMXB
formation (see also \S 4.4 and 5).

\subsection{X-ray Properties of GC$-$ and Field$-$LMXB}

\begin{figure}
\epsfig{figure=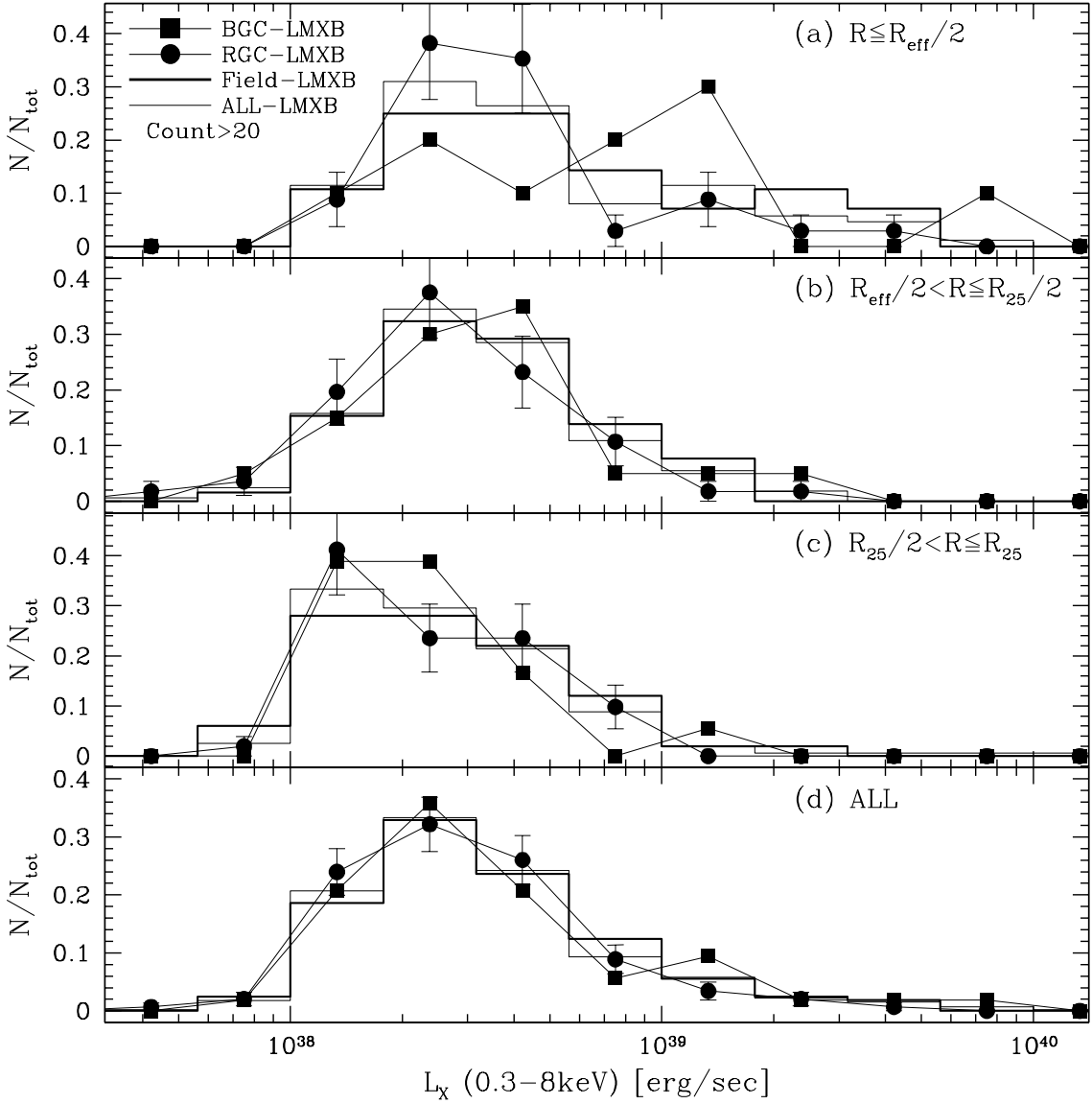, height=0.975\textwidth, width=0.975\textwidth}
\vspace{-3mm}
\caption{Differential luminosity functions
of X-ray point sources for
(a) the central region (b) the intermediate region (c) the outer regions, and
(d) the all radial regions.
The filled squares, filled circles, thick solid line and the thin solid lines
represent BGC$-$LMXBs, RGC$-$LMXBs, field$-$LMXBs and
the whole LMXB sample, respectively.}
\label{fig-rxlum}
\end{figure}

The X-ray luminosity distributions (normalized to the total number of
sources) of LMXBs in different radial regions are shown in
Fig. \ref{fig-rxlum}.  We only
consider the X-ray point sources with net counts $> 20$ to minimize
the incompleteness.  The X-ray luminosity distributions of different
LMXB populations are statistically indistinguishable. The KS tests for
any combination of two samples in the same radial bin suggest that
the hypothesis that they are drawn from different populations is
excluded at the $95\%$ confidence limit. The only possible exception is
the BGC$-$LMXB sample of the central region which appears to peak at
higher $L_X$. However this difference may be due to a small number
(11) of BGC$-$LMXBs.

%

\begin{figure}
\vspace{-20mm}
\hspace{-10mm}
\epsfig{figure=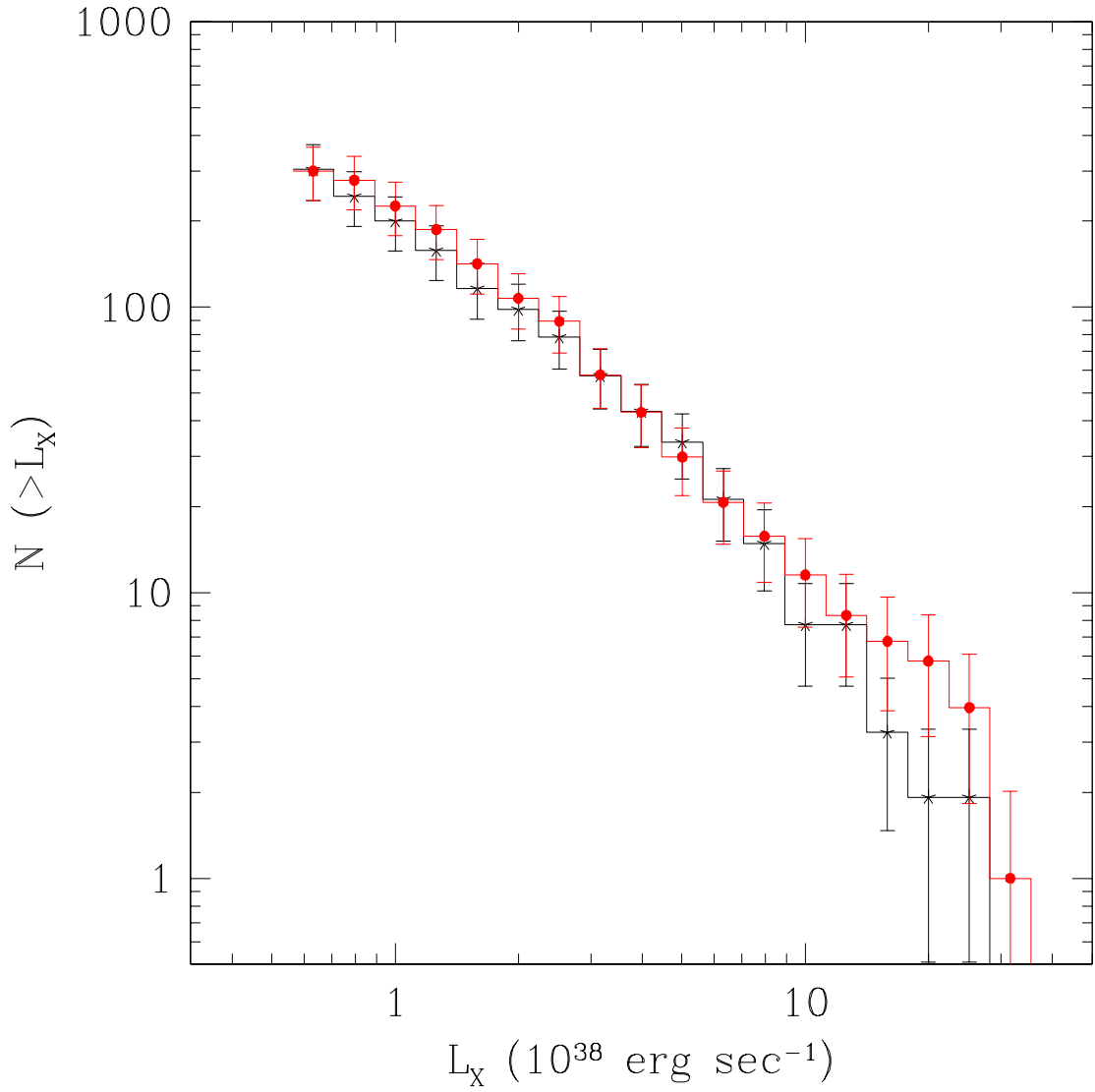, height=1.600\textwidth, width=1.150\textwidth}
\vspace{-60mm}
\caption{XLFs of field$-$LMXBs (asterisks) and GC$-$LMXBs (filled circles)
in six elliptical galaxies.}
\label{fig-xlf}
\end{figure}

Fig. \ref{fig-xlf} shows completeness-corrected X-ray luminosity functions
(XLFs) of GC$-$LMXBs and field$-$LMXBs, derived
by applying the simulation technique as described in
\citet{kf04}. They are statistically indistinguishable in their shape.
Their normalizations are also identical within the error, but this
result is not astrophysically meaningful, because of galaxy to
galaxy variations as noted in \S 3.2 (see also \S 5).  The
only possible difference is observed at the most luminous end ($L_X > 10^{39}$
erg s$^{-1}$) of the XLFs, where there are more GC$-$LMXBs than field$-$LMXBs, but
with a limited significance ($\sim$ 1 $\sigma$) due to the small
number of very bright LMXBs. We will discuss the implications of this
result for the presence of BH X-ray binaries in GCs in \S 5.  The
individual XLFs of BGC$-$LMXBs and RGC$-$LMXB are also consistent
with each other, except that the XLF of BGC$-$LMXBs is slightly (but within a
$1\sigma$ error) flatter. We fit the observed XLFs with both single
power-law and broken power-law models and find that our results are
consistent with those of 14 early type galaxies studied by
\citet{kf04}. The best-fit slope for a single power-law is
$2.1\pm0.13$ for both GC$-$ and field$-$LMXBs with $\chi^2_{red}$
close to $1$. Since the fit with a single power law is
already good and a more complex model is not required, the parameters
of the broken power law are not well constrained, but our results
($L_{break} = 2-10 \times 10^{38}$ erg s$^{-1}$; slope = $1.5-2.0$ and
$2.0-5.0$ below and above the break)  are consistent with
\citet{kf04}.

\begin{figure}
\epsfig{figure=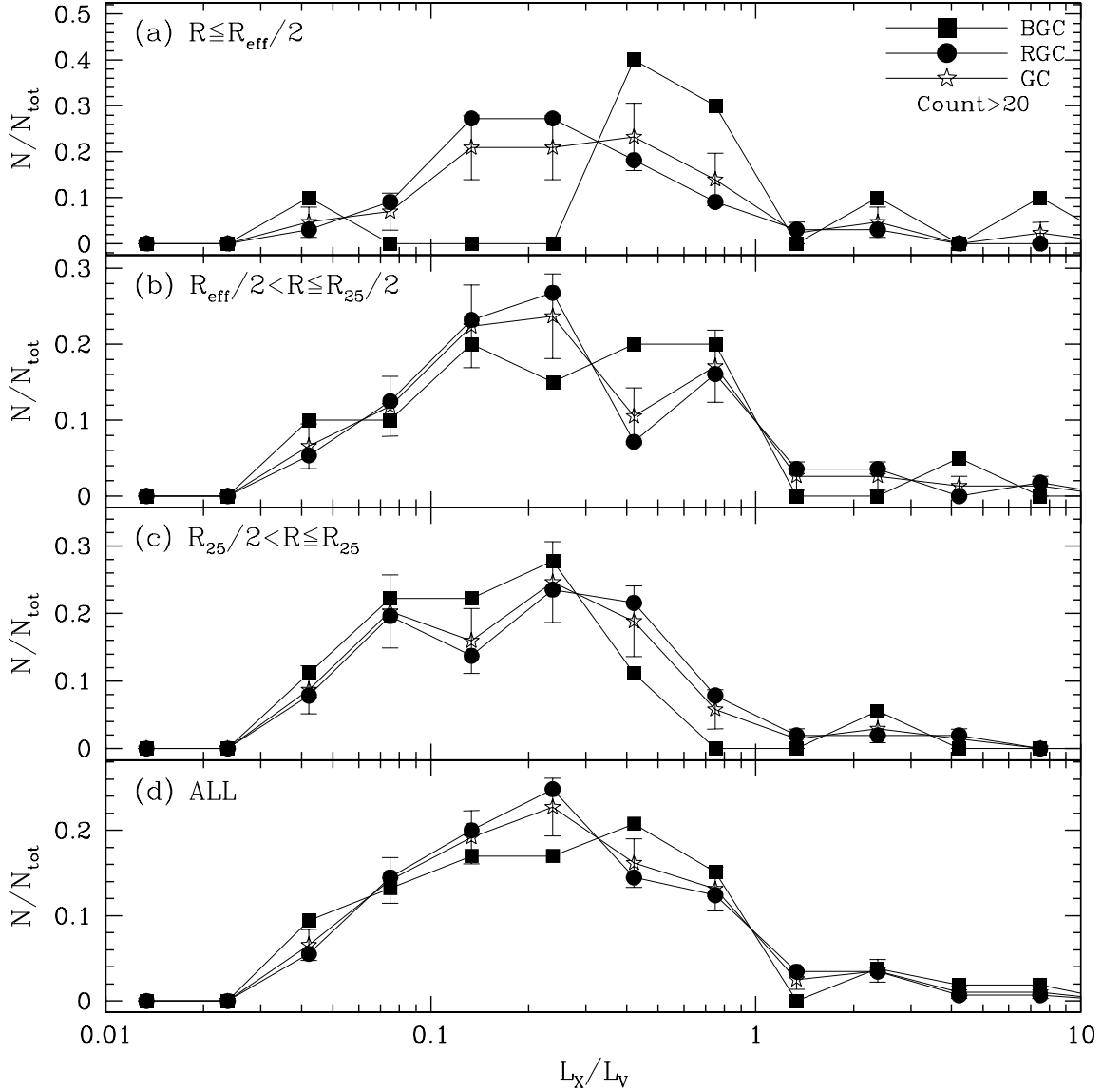, height=0.975\textwidth, width=0.975\textwidth}
\vspace{-3mm}
\caption{Distribution of $L_X/L_V$ for
(a) the central region (b) the intermediate region (c) the outer regions and
(d) the all radial regions.}
\label{fig-rxolum}
\end{figure}

The $L_X/L_V$ luminosity ratios for LMXBs found in GCs are displayed
in Fig. \ref{fig-rxolum}. Again, there is no statistically significant
difference in $L_X/L_V$ between any combination of two
sub-samples. Again the BGC$-$LMXB sample of the central region peaks
at the higher $L_X/L_V$, but this effect is not conclusive because of
the large errors. Note that a typical LMXB X-ray luminosity is
roughly $10-30\%$ of the optical luminosity in $V$ band of an entire
GC, and a few LMXBs are more luminous than their host GCs ($L_X/L_V >
1$).


To investigate the X-ray spectral properties of different sub-samples,
we use the X-ray hardness ratio, defined as HR = (H$-$S) / (H$+$S),
where S and H are net counts in $0.5 - 2.0$ keV and $2.0 - 8.0$ keV,
respectively.
We also use X-ray colors as defined in \citet{kimetal04},
C21 = log (C1/C2) and C32 = log (C2/C3),
where C1, C2 and C3 are net counts in $0.3 - 0.9$ keV,
$0.9 - 2.5$ keV, and
$2.5 - 8.0$ keV, respectively. By definition, as the X-ray spectra become
harder, HR increases and X-ray colors decrease. 
For faint sources with a small number of counts, HR and colors often
result in unrealistic values with unreliable errors because of
negative net counts in one band and a non-symmetric Poisson
distribution. We apply a Bayesian approach developed by \citet{dyk04},
which models the detected counts as a non-homogeneous
Poisson process.
Taking into account the ACIS QE degradation which could
change the soft band counts by $<20\%$ \citep{kimetal04}, we also
convert the counts to what would be obtained at the mid-point within
the observation period of our sample.
The mean and standard deviations of each group are listed in Table
\ref{tab-xprop}. We find no
statistically significant differences in the X-ray HRs and colors of
field$-$LMXBs and GC$-$LMXBs and also between
RGC$-$LMXBs and BGC$-$LMXBs.


\subsection{Radial Profiles}

We explored spatial differences by comparing
the radial profiles of the surface number density of LMXBs and
GCs.
Because the radial distribution of optical GCs
is known to be flatter than that of the optical halo light
\citep[e.g.,][]{lk00},
it is particularly interesting to test whether the
radial profiles of GC$-$LMXBs and field$-$LMXBs follow that of the optical
GCs or that of the optical halo light.

\begin{figure}
\epsfig{figure=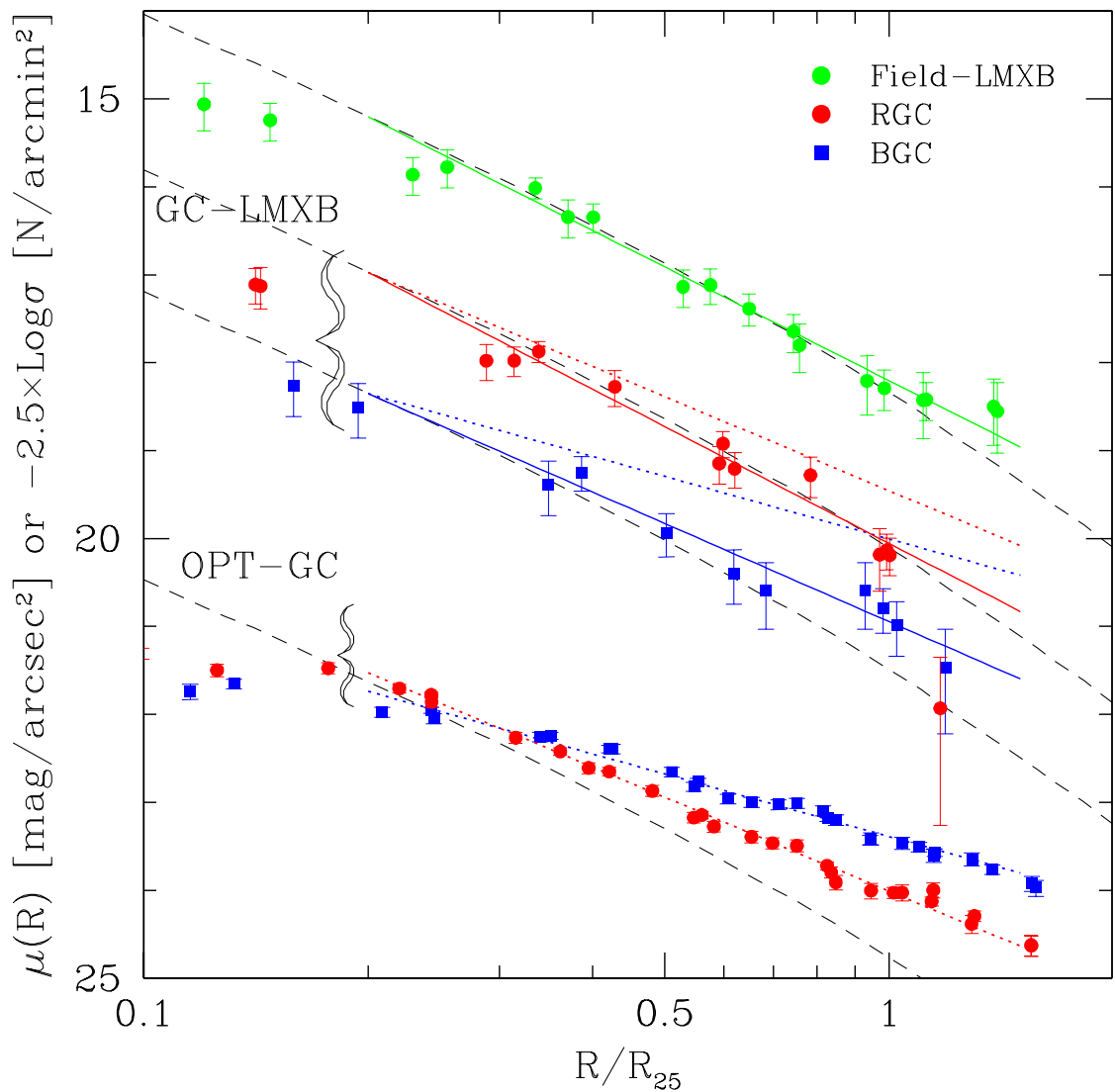, height=0.975\textwidth, width=0.975\textwidth}
\vspace{-3mm}
\caption{Radial profiles of point sources and the combined galaxy light.
The radial profiles for optical
GCs in the bottom part and LMXBs in the top part are scaled to be 
compared directly to that of the
optical halo light. The curved dashed lines represents the scaled galaxy
halo light after
shifting upward and downward for easy comparison with other profies.
The different symbols represent BGC (blue squares),
RGC (red circle) and field$-$LMXBs (green circle).
The red/blue solid line shows best-fit for LMXBs in RGC/BGC while
the red/blue dotted line for optical RGC/BGC.}
\label{fig-rprofall}
\end{figure}

We show the combined radial profiles of the surface
density of LMXBs, optical GCs and galaxy halo light (Fig. \ref{fig-rprofall}).
The linear least square fit for the
radial range of $0.2 < R/R_{25} < 1.2$ and the slopes of the fit are
summarized in Table \ref{tab-rprof2}. The minimum radius is chosen to
minimize missing sources due to highly varying background in both optical
and X-ray images.

As seen in the bottom part of Fig. \ref{fig-rprofall}, the radial
profile of GCs is considerably flatter than that of the optical halo
light. This trend is more pronounced in BGCs than in RGCs.  This
result is consistent with previous optical studies
\citep[e.g.,][]{lk00}.  However, surprisingly, we find that the radial
profile of GCs with LMXBs is significantly steeper than that of the
optical GC population, but close to that of the halo light (see the
top part of Fig. \ref{fig-rprofall}).  The difference of the radial
profile slope between the whole GCs and GCs with LMXBs is 1.45, which
corresponds to a statistical significance of $> 6\sigma$ (see Table
\ref{tab-rprof2}).  This result is consistent with our result (\S
4.2) of GCs having a higher chance to harbor LMXBs in the central
region and suggests an important clue for the LMXB formation mechanism
(see \S 5).  This trend is also valid for RGCs
and BGCs separately.  While the radial profile of RGCs (either the
entire GC sample or GCs with LMXBs) is steeper than that of BGCs, the
radial profile of GCs with LMXBs is still steeper than the optical GCs
in both BGCs and RGCs samples.  It is also interesting to note that
the radial profile slope of the field$-$LMXBs is very similar to that
of RGC$-$LMXBs, while steeper than, but still consistent with that of
BGC$-$LMXBs.

\begin{figure}
\epsfig{figure=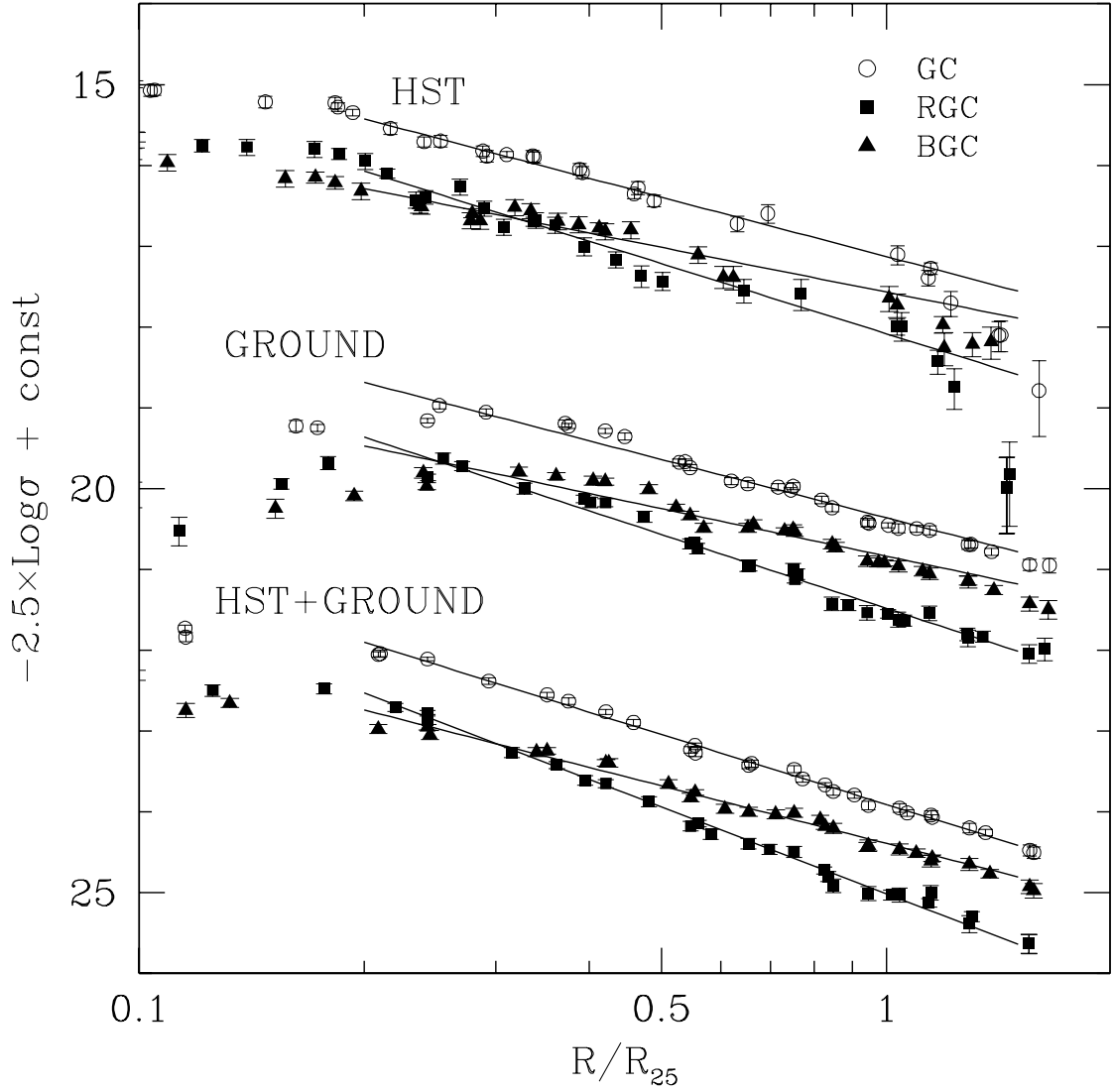, height=0.975\textwidth, width=0.975\textwidth}
\vspace{-3mm}
\caption{Radial profiles of optical globular cluster candidates for
(1) {\hst} observation
(2) ground-based observation
(3) combined list of ground and {\hst} observations.}
\label{fig-rprofopt}
\end{figure}

To make sure that these differences in radial distributions are not a
statistical fluke, we performed 1000 simulations by randomly selecting
the same number of GCs as observed from the whole GC sample, then
fitting the radial profile by the same method which used in
Fig. \ref{fig-rprofall}. We find that the probability that the random
sample has a steeper slope than the observed is $0.4\%$ ($2.8\%$ and
$8.7\%$ for BGC and RGC separately). To further check for possible
contamination by foreground stars and background galaxies, we compared
the radial profiles produced by the {\hst} data and the ground optical
observations separately (Fig. \ref{fig-rprofopt}).
 As demonstrated in \S 2, the
contamination in our sample of optical GCs is small based on
comparison between the {\hst} and ground-based data, the GC radial
profiles produced by ground-based and {\hst} data are consistent with
each other within a $1\sigma$ error for all sub populations of GCs.

\section{Discussion}

Using {\chan} and Hubble archival data of 6 elliptical galaxies,
supplemented by deep optical ground-based imaging observations, we
identify 6173 GCs and 665 LMXBs within the $D_{25}$ ellipse of these
galaxies (\S 2). Applying conservative matching criteria, we find
285 LMXBs coincident with GCs (209 in RGC and 76 in BGC) and 259 LMXBs
in the field (\S 3). This is the largest GC $+$ LMXB sample studied
so far.

\subsection{Metallicity and LMXB formation in GCs}

We find that the probability to find LMXBs is on average $\sim$ 3
times higher for RGCs than BGCs, but with a non-negligible variation
from one galaxy to another (\S 3.2), consistent with previous
reports of early type galaxies \citep[e.g.,][]{kmz02,sarazin03,jordan04},
, and in the Milky Way and M31
\citep[e.g.,][]{grind93}. This result indicates that metal abundance plays a
key role in forming LMXBs in the globular clusters, as suggested by
the above authors.

The physical mechanism linking the metallicity and the formation and
evolution of LMXBs is not well understood.
\citet{belletal95} suggested that the larger stellar size of a
metal rich star can increase the tidal capture rate, making it easier
to fill the Roche-lobe, and therefore may be responsible for the
preferential association of LMXBs with metal rich RGCs.  Instead,
\citet{mkz04} showed that the effect of the larger stellar
size is not enough to explain the observed difference, and proposed an
irradiation induced stellar wind model, where a metal-poor star (in
BGC) with a stronger stellar wind evolves more rapidly than a
metal-rich star in RGC.  This model predicts harder X-ray spectra in
BGC$-$LMXBs than RGC$-$LMXBs, because of the extra absorption by accreting
materials in BGC$-$LMXBs (their estimated column density $N_H \sim 6
\times 10^{21}$ cm$^{-2}$). 

However, when we compare the X-ray spectral hardness/absorption of RGC
and BGC sources (\S 4.3), we find no statistically significant
differences.  Although we cannot rule out a small amount of intrinsic
absorption given the statistical uncertainty, we estimate that $N_H$
cannot exceed $10^{21}$ cm$^{-2}$ in both RGCs and BGCs.  Therefore,
our result does not support the prediction of the stellar wind
model. Recently, Ivanova (2005) suggests that the absence of an outer
convective zone in the metal poor main sequence star may explain the
observed trend. Because magnetic breaking, necessary for the orbital
decay to form a compact X-ray binary, does not turn on without the
outer convective zone, the efficiency to form LMXBs is
considerably lower in BGCs than in RGCs. If this scenario is correct,
there is no specific reason for different X-ray obscuration and the
X-ray spectral properties of LMXBs in RGCs and BGCs would be similar,
as indicated in our results. However, no theoretical model, so far,
can explain why the fraction of LMXBs in RGCs and BGCs considerably
varies from one galaxy to another (from 1.4 to 4.6; \S 3.2).

\subsection{Dynamical Effect in GC-LMXB formation}

We have found that the GCs located near the center of galaxies have a
higher probability to harbor LMXBs compared to those in the
outskirts. This trend is confirmed with a high significance both by
the luminosity dependent GC$-$LMXB fractions in different regions
(\S 4.2) and by the radial profiles of LMXBs and GCs (\S
4.4). Although a negative radial gradient of the average GC
metallicity is known in some elliptical galaxies \citep[e.g.,][]{lkg98}
, it can be mostly attributed to the steeper radial profile
of metal-rich RGCs, compared to that of metal-poor BGCs
\citep[][as seen in Fig. \ref{fig-rprofall}]{glk96}
Therefore, the metallicity gradient for the RGC and BGC
individually is not significant \citep{lkg98} and cannot explain
the observed higher probability to harbor an LMXB near the galactic
center than in the outskirts. A secondary mechanism, dependent on the
galacto-centric distance, must play an important role in the GC$-$LMXB
formation, with the metallicity being the primary mechanism as
discussed above. One possible explanation is that GCs near the
galactic center may have a compact core and a higher central density
than GCs in the outer regions, as a result of selective GC disruption
by the galactic tidal force. This conclusion is also consistent with
the recent {\hst} study of M31 where the central density of GCs increases
toward the center of M31 (Barmby, Holland, \& Huchra 2002; see also
Bellazzini et al. 1995). Therefore, the chance to form LMXBs is
expected to increase in GCs near the galaxy center because of a
higher rate of either tidal capture or exchange interaction resulting
from the higher central stellar density.

Based on the structural parameters of individual GCs determined with the
{\hstacs} data of M87, Jordan et al. (2004) showed that the encounter
rate ($\Gamma$) of GCs with LMXBs is considerably higher than the mean
$\Gamma$. They formulated a probability of a GC hosting an LMXB as a
function of $\Gamma$ and Z. This formulation is consistent with
similar results on Galactic GCs by \citet{pooley03} and
\citet{heinke03}, and with the higher central
density of M31 GCs with LMXBs (Bellazzini et al. 1995). Our results
further support that the probability of harboring an LMXB requires a
secondary parameter in addition to metallicity.

Based on the correlation between $\Gamma$ and luminosity, Jordan et
al. (2004) also suggested that the luminosity dependency on the
fraction of GCs with LMXBs may be a consequence of a more fundamental
relation between $\Gamma$ and luminosity.  As reported previously
(e.g., Sarazin et al. 2003), we confirm that the more luminous GCs
have a higher probability to host LMXBs (\S 4.1). This trend is
valid for each RGC and BGC subsample.  However, we find that the
linearity holds only in RGCs, but not in BGCs. The RGC$-$LMXB fraction
increases by a factor of $\sim$ 5.5 as $M_V$ increases from --8 mag to
--9.8 mag (i.e., brighter by a factor of 5.2). This is consistent with
the expected linear increase of the LMXB fraction with increasing
luminosity.  On the other hand, the BGC$-$LMXB fraction increases only
by a factor of $\sim$ 2 with the same optical magnitude range. This
may indicate a complex relation in metal poor GCs between the cluster
luminosity (or mass) and the LMXB fraction, as might be suggested by
Ivanova (2005).


\subsection{Can BH X-ray binaries form in GCs?}

It is well known that no BH X-ray binary has been found in Galactic
GCs \citep[e.g.,][]{grindetal01}; however, the total number of GC$-$LMXBs
in the Milky Way is relatively small. \citet{kalogera04} showed that
the duty cycle for a BH binary formed in the center of a dense cluster
by an exchange interaction is extremely low. For early type galaxies,
mixed results have been reported. \citet{sarazin03} reported a weak
tendency for bright LMXBs to avoid GCs and
\citet{miniti04} reported that no bright GC$-$LMXB (ie., BH candidates)
is identified in NGC 5128. On the other hand, a number of luminous GC$-$
LMXBs are found in N1399 \citep{alm01} and M87 \citep{jordan04}.
\citet{alm01} have even claimed that GC$-$LMXBs are on the average
more luminous than the field LMXBs. We found with high statistical
confidence (\S 4.3) that the XLFs of GC$-$LMXBs and field$-$LMXBs are
statistically consistent, indicating that luminous LMXBs are equally
found in GCs and in the field. More specifically, there are 26
GC$-$LMXBs and 27 field LMXBs with $L_X > 5
\times 10^{38}$ erg sec$^{-1}$.  We note that $L_X = 5 \times 10^{38}$
erg sec$^{-1}$ corresponds to the break luminosity of the LMXB XLF
where NS and BH X-ray binaries are likely separated, as identified by
\citet{kf04}. More conservatively, if we consider $L_X > 10^{39}$ erg
sec$^{-1}$, there are still 8 GC$-$LMXBs and 3 field$-$LMXBs.
Therefore, our results do not support the hypothesis
that a GC cannot harbor a BH X-ray binary.

\subsection {Are field LMXBs formed in the field or in GCs?}

We find on average an equal number of LMXBs in GCs and in the field,
although with a non-negligible galaxy to galaxy variation. One of the
key questions to understand LMXB formation is whether GCs are the only
birth place for all LMXBs.  In this scenario, field$-$LMXBs were
originally formed in GCs and they were then ejected from the parent GC
or left alone in the field as the GC was disrupted.  Utilizing the
fact that the radial profile of GCs is flatter than that of the
galactic halo light \citep[e.g.,][]{lkg98}, we tested this hypothesis
by determining whether the radial profile of field$-$LMXBs follows that
of GCs or that of the halo light.  We found (\S 4.4) that regardless of
their association with GCs, LMXBs are distributed like the optical halo
light, not the GCs.  The close agreement between the radial
distributions of the optical light and LMXBs has also been reported in NGC
1316 \citep{kf03}, NGC 1332 \citep{hb04} and NGC 4486
\citep{jordan04}. This seems to suggest a rather complex connection, depending on
various factors operating in the LMXB formation in GCs and its
subsequent evolution (see \S 5.2). Because of the similar radial
distribution between GC$-$LMXBs and field$-$LMXBs, we can neither prove nor
reject the hypothesis whether field$-$LMXBs were originally formed in
the GC.

If field$-$LMXBs were ejected from GCs, they may still be in the
neighbourhood of the parent cluster. We therefore searched for a
nearby GC, which could have been the previous host of the current
field$-$LMXB, and compared the results with the expectation from the
mean space density of GCs. Firstly, we find that the mean angular
distance of the nearest GC from the field LMXBs is $5-10$ arcsec. This
is compatible with the expected mean random separation, based on the
space density of GCs ($\sim$ 8 arcsec), indicating that there is no
preference to find a potential host GC near a field-LMXB.  Secondly,
we measure the GC surface density near the field-LMXBs.  Again, the
estimated density is comparable with that expected from the GC space
density ($\sim 18$ GCs/arcmin$^2$).

\begin{figure}
\epsfig{figure=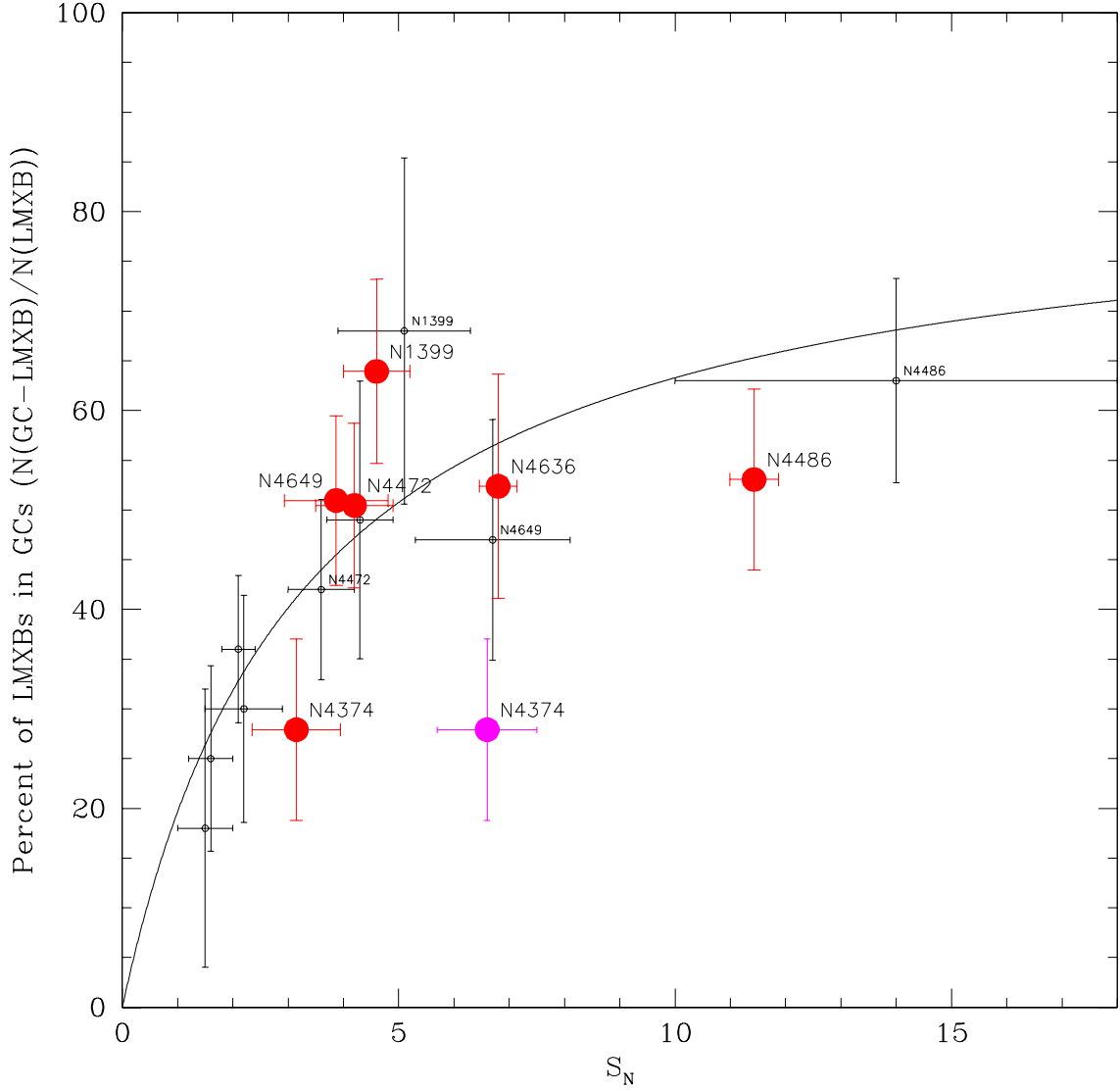, height=0.975\textwidth, width=0.975\textwidth}
\vspace{-3mm}
\caption{Fraction of the LMXB population found in GCs plotted against the
GC-specific frequency, $S_N$. The data for galaxies in the present study are
shown with red circles while
the data from \citet{juett05} with small open circles.
Two values of $S_N$ for NGC 4374 are from \citet{gr04} (red) and \citet{kissler97} (magenta).}
\label{fig-sn}
\end{figure}

Based on simple relations of $f$(GC$-$LMXB) and $L_X$(LMXB) against
the GC specific frequency ($S_N$), \citet{juett05} and \citet{irwin05}
suggested that a considerable fraction of field$-$LMXBs were indeed
formed in the field. Although this suggestion is intriguing,
reality may be more
complex. Their relations may test the hypothesis of field$-$LMXBs
ejected from the parent GCs, but cannot work if the parent GCs were
disrupted, because the current $S_N$ would not include GCs disrupted
in the past. Furthermore, because $S_N$ is usually determined in a
large scale (e.g., compared to the {\hst} field of view), it is important
to determine LMXB properties at a comparable scale to compare with
$S_N$. However, the previous studies are mostly limited to the {\hst}
field of view. If we plot our data of Table 10 in the same figure of
\citet{juett05}, the relation appears to be less convincing (Fig. \ref{fig-sn}).
Most galaxies (4 out of 6) have almost constant
$f$(GC$-$LMXB) (50 $\pm$ 5\%), but with a wide range of $S_N = 4 - 12$
\citep[from][]{dirsch05,kissler97,rz01,forbesetal04}.
The remaining two galaxies are also
outliers. NGC 4374 has the smallest GC$-$LMXB fraction (30\%) in our
sample, but with a rather modest $S_N$ (3.2, taken from G\'{o}mez \& Richtler
2004; Harris 1991 or 6.6 from Kissler-Patig 1997).
NGC 1399 has the highest GC$-$LMXB fraction (65\%) in our sample,
but $S_N$ is only 4.6 \citep{dirsch03}. Therefore, the
proposed relationship with $S_N$ does not appear to be
straightforward. The situation could be even more complex, for
example, because of different merger histories (e.g., Schweitzer 2003)
and different degrees of environment-dependent GC stripping (Bekki et
al. 2003) which would add significant galaxy to galaxy variations.
In particular, we note that three galaxies with the lowest GC$-$LMXB
fraction (NGC 1553, NGC 3115, NGC 1332) are all S0 galaxies.






\section{Conclusions}

1. In our sample of six elliptical galaxies, we find 285 LMXBs matched
with GCs (209 in RGC and 76 in BGC) and 259 LMXBs in the field. This
is the largest sample studied so far. We estimate that the systematic
error in LMXB $-$ GC associations due to the source contamination and
false matches is $5-10\%$.

2. We confirm that on average the fraction of RGCs with LMXBs is three
times higher that that of BGCs with large variations from one galaxy
to another, indicating that metallicity is an important factor in GC$-$LMXB
formation \citep{(kmz02,sarazin03,jordan04}.
We find that the average X-ray spectra of RGC$-$LMXBs and BGC$-$LMXB
are statistically identical, in disagreement with the prediction
of the stellar wind model \citep{mkz04}, but consistent with
the explanation of the lack of outer convective zone in BGCs (Ivanova
2005). We also find that while the brighter (and bigger) GCs have a
higher probability to host LMXBs as suggested by Sarazin et
al. (2003), this linear dependency on the optical luminosity only
holds in RGCs, possibly implying a complex formation scenario in
BGCs (e.g., Ivanova 2005).

3. Both RGCs and BGCs located near the galaxy center have a higher
probability to harbor LMXBs compared to GCs at the outer radii.  The
same trend is also confirmed by the steeper radial profile of
GC$-$LMXBs (for both RGC and BGC), when compared to that of the whole
GC sample. This suggests there must be another parameter (in addition
to metallicity) for LMXB formation in GCs, which critically depends on
the galacto-centric distance. One possibility is a variable encounter
rate, depending on the galacto-centric distance, as suggested by
Jordan et al. (2004).

4. We find no statistically significant difference in the X-ray
properties (shape of X-ray luminosity function, $L_X/L_V$
distribution, X-ray spectra) among RGC$-$LMXBs, BGC$-$LMXBs and
field$-$LMXBs. In particular, there is no observational preference to
host or avoid BH X-ray binaries in GCs.

5. We find on average an equal number of LMXBs in GCs and in the
field. We have tested the hypothesis that field$-$LMXBs were once formed
in GCs, by comparing radial profiles of GC$-$LMXBs and field$-$LMXBs and
by searching for possible parent GCs near field$-$LMXBs. We find that
LMXBs, regardless of their association with GCs, do not follow the
radial distribution of GCs, but more closely follow that of the
optical halo light. The average distance and density of GCs near the
field$-$LMXBs are consistent with the expectation from the mean GC space
density. Therefore, we could not prove or reject this hypothesis.





\acknowledgments

This work was supported by Chandra GO program GO3-4109X and
Chandra archival research grant AR7-0001X.
DWK and GF acknowledge support through NASA contract NAS8-39073 (CXC).
MGL was supported in part by grant R01-2004-000-10490-0 from the 
Basic Research Program
of the Korea Science and Engineering Foundation.

\clearpage


\begin{table}
\centering
\begin{minipage}{0.80\textwidth}
\caption{List of target galaxies\label{tab-target}}
\begin{tabular}{@{}ccccccccc@{}} \hline\hline
Name     & R.A.        & Dec         & $B_T^1$ & $(m-M)_0^2$ & $\epsilon^1$ & P.A.[$^o$]$^1$\\ \hline\hline
NGC 1399 & 03 38 29.32 & $-$35 27 00.7 & 10.55 & 31.4 & 0.07 &   0 \\
NGC 4374 & 12 25 03.74 & $+$12 53 13.1 & 10.09 & 31.2 & 0.13 & 135 \\
NGC 4472 & 12 29:46.76 & $+$07 59:59.9 &  9.37 & 31.2 & 0.18 & 155 \\
NGC 4486 & 12 30 49.42 & $+$12 23 28.0 &  9.59 & 31.2 & 0.21 &   0 \\
NGC 4636 & 12 42 50.00 & $+$02 41 16.5 & 10.43 & 31.2 & 0.22 & 150 \\
NGC 4649 & 12 43 40.19 & $+$11 33 08.9 &  9.81 & 31.2 & 0.19 & 105 \\ \hline
\end{tabular}

\vspace{2truemm}
$^1$ Data from the RC3 catalog \citep{deV91}

$^2$ NGC 1399 \citep{ffg05} \& \citet{lkg98} for the other galaxies
\end{minipage}
\end{table}

\clearpage


\begin{table}
\centering
\begin{minipage}{0.70\textwidth}
\caption{Archival {\chan} ACIS observations\label{tab-xobs}}
\begin{tabular}{@{}cccc@{}} \hline\hline
Name & Obsid & Observation Date & Exposure$^1$ \\ \hline\hline
NGC 1399 &  319 & 2000-01-18 &  56 \\
NGC 4374 &  803 & 2000-03-19 &  27 \\
NGC 4472 &  321 & 2000-06-12 &  34 \\
NGC 4486 & 2707 & 2002-07-06 & 100 \\
NGC 4636 &  323 & 2000-01-26 &  42 \\
NGC 4649 &  785 & 2000-04-20 &  22 \\ \hline
\end{tabular}

\vspace{2truemm}
$^1$ effective exposure times in unit of ksec
\end{minipage}
\end{table}
\clearpage


\begin{table}
\centering
\begin{minipage}{0.70\textwidth}
\caption{Journal of optical ground-based observations\label{tab-gobs}}
\begin{tabular}{@{}ccccc@{}} \hline\hline
Name & Observation Date & Telescope & Filters & references \\ \hline\hline
NGC 1399 & 1999-12-07 & CTIO $4m$ &  $CR$ & 1 \\
NGC 4374 & 1997-04-08 & KPNO $4m$ &  $CR$ & 2 \\
NGC 4472 & 1993-02-26 & KPNO $4m$ &  $CT_1$ & 3,4 \\
NGC 4486 & 1997-04-09 & KPNO $4m$ &  $CR$ & 2 \\
NGC 4636 & 1997-04-10 & KPNO $4m$ &  $CR$ & 2 \\
NGC 4649 & 1997-04-09 & KPNO $4m$ &  $CR$ & 2 \\ \hline
\end{tabular}

\vspace{2truemm}

references: (1) \citet{dirsch03}, (2) \citet{lee05}, (3) \citet{glk96}, (4) \citet{lkg98}

\end{minipage}
\end{table}
\clearpage


\begin{table}
\centering
\begin{minipage}{0.70\textwidth}
\caption{Journal of {\hstwfpc2} observations\label{tab-hobs}}
\begin{tabular}{@{}ccccc@{}} \hline\hline
Name & Field & \multicolumn{2}{c}{T$_{exp}$ [sec]} & Program ID \\ \cline{3-4}
         &    &      $V^1$      &   $I^2$ &      \\ \hline\hline
NGC 1399 & C1 & $-$             & $1,800$ & 5990 \\
         & C2 & $4,000(F606W)$  & $-$     & 8214 \\
         & N  & $3,500(F606W)$  & $900$   & 9244 \\
         & S  & $460$           & $300$   & 6352 \\[1pt] \hline
NGC 4374 & C  & $280(F547M)$    & $520$   & 6094 \\[1pt] \hline
NGC 4472 & C1 & $1,800$         & $1,800$ & 5236 \\
         & C2 & $520$           & $520$   & 5236 \\
         & N  & $2,200$         & $2,300$ & 5920 \\ 
         & S  & $2,200$         & $2,300$ & 5920 \\[1pt] \hline
NGC 4486 & C  & $2,430$         & $2,430$ & 5477 \\
         & N1 & $2,000$         & $1,800$ & 6844 \\
         & N2 & $400$           & $800$   & 7274 \\
         & S  & $2,000$         & $1,800$ & 6844 \\
NGC 4636 & C  & $1,000(F547M)$  & $400$   & 8686 \\
         & S  & $1,800$         & $1,820$ & 8686 \\[1pt] \hline
NGC 4649 & C  & $2,100$         & $2,500$ & 6286 \\
         & N  & $4,800$         & $9,600$ & 7388 \\ \hline
\end{tabular}

\vspace{2truemm}

$^1$ $F555W$ if not mentioned

$^2$ $F814W$

\end{minipage}
\end{table}

\clearpage


\begin{table}
\footnotesize
\centering
\begin{minipage}{1.00\textwidth}
\caption{Structural parameters of the studied elliptical galaxies\label{tab-sphot}}
\begin{tabular}{@{}c|c|c|c|c|c|c@{}} \hline\hline
Parameter &           NGC 4374 & NGC 4486 & NGC 4636 & NGC 4649 & NGC 4472$^2$ & NGC 1399$^3$ \\ \hline \hline
$R_{eff} [']$          &  1.20 &  1.53 &  1.49 &  1.50 & 2.00  & 2.50  \\
$\epsilon_{eff}$       & 0.065 & 0.125 & 0.256 & 0.216 & 0.175 & 0.099 \\
PA$_{eff} [^o]$        &    67 &   159 &   148 &   112 & 155   & $-$   \\
$(C-R)_{eff}$          &  1.87 &  1.86 &  1.69 &  1.88 & 1.83  & $-$  \\
$R_{25} [']$           &  3.63 &  4.36 &  3.60 &  4.03 & 5.22  & 3.93  \\
$\epsilon_{25}$        & 0.016 & 0.320 & 0.326 & 0.224 & 0.200 & 0.17  \\
PA$_{25}  [^o]$        &    67 &   159 &   148 &   112 & 155   & $-$   \\
$(C-R)_{25}$           &  1.74 &  1.93 &  1.45 &  1.92 & 1.88  & $-$  \\
$<C-R>^1$              &  1.81 &  1.88 &  1.57 &  1.85 & $-$   & $-$   \\
$\Delta\mu(R)/\Delta$logR=1$^1$ & 4.64 &  4.60 &  5.20 & 5.13  & $-$ & \\
$\Delta(C-R)/\Delta$logR=1$^1$ & $-$0.232  & 0.219 & $-$0.612 & 0.036 & $-$0.08 & $-$ \\ \hline \hline
\end{tabular}

\vspace{2truemm}
$^1$ values computed between effective radius and standard radius

$^2$ \citet{klg00}

$^3$ \citet{dirsch03}

\end{minipage}
\end{table}

\clearpage


\begin{table}
\centering
\begin{minipage}{0.80\textwidth}
\caption{Adopted color boundaries of GCs\label{tab-gcbox}}
\begin{tabular}{@{}cccccc@{}} \hline\hline
Name & \multicolumn{2}{c}{$C-R$}   & & \multicolumn{2}{c}{$V-I$}\\ \cline{2-3}\cline{5-6}
     &        BGC     &    RGC     & &    BGC     &    RGC     \\ \hline\hline
NGC 1399 & 1.00 .. 1.65 & 1.65 .. 2.20 & & 0.65 .. 1.00 & 1.00 .. 1.45 \\
NGC 4374 & 1.00 .. 1.55 & 1.55 .. 2.20 & & 0.70 .. 1.08 & 1.08 .. 1.40 \\
NGC 4472 & 1.00 .. 1.65 & 1.65 .. 2.20 & & 0.75 .. 1.08 & 1.08 .. 1.45 \\
NGC 4486 & 0.90 .. 1.65 & 1.65 .. 2.20 & & 0.80 .. 1.13 & 1.13 .. 1.45 \\
NGC 4636 & 0.90 .. 1.50 & 1.50 .. 2.10 & & 0.75 .. 1.15 & 1.15 .. 1.50 \\
NGC 4649 & 1.00 .. 1.65 & 1.65 .. 2.20 & & 0.80 .. 1.15 & 1.15 .. 1.50 \\ \hline
\end{tabular}

\vspace{2truemm}

\end{minipage}
\end{table}

\clearpage


\begin{table}
\centering
\begin{minipage}{0.85\textwidth}
\caption{Optical GCs\label{tab-psrc1}}
\begin{tabular}{@{}cccccccccc@{}} \hline\hline
Region & Name & \multicolumn{2}{c}{BGC} & & \multicolumn{2}{c}{RGC} & & \multicolumn{2}{c}{Contamination} \\ \cline{3-4} \cline{6-7} \cline{9-10}
       &      &  G$^1$ & H$^2$          & &  G$^1$ & H$^2$          & &    BGC & RGC \\ \hline\hline
                       & NGC 1399 & 134 &  72 & & 153 & 148 & & 25 & 38 \\
                       & NGC 4374 &  20 &  48 & &  21 &  42 & &  3 &  2 \\
                       & NGC 4472 &  33 & 178 & & 32 & 244 & &  3  & 2 \\
$20'' \le R < R_{eff}$ & NGC 4486 &  87 & 245 & & 72 & 278 & & 13 & 10 \\
                       & NGC 4636 &  46 & 123 & & 65 &  95 & &  7 & 10 \\
                       & NGC 4649 &  38 & 102 & & 43 & 121 & &  3 &  4 \\ \cline{2-10}
                       &   total  & 358 & 768 & & 386& 928 & & 54 & 66 \\ \hline
                         & NGC 1399 & 154 &   5 & &  129 &   4 & &  28 &  32 \\
                         & NGC 4374 & 207 &  23 & &  142 &  20 & &  28 &  14 \\
                         & NGC 4472 & 341 &  58 & &  194 &  53 & &  27 &  12 \\
$R_{eff} \le R < R_{25}$ & NGC 4486 & 680 & 154 & &  287 & 103 & & 103 &  40 \\
                         & NGC 4636 & 265 &  19 & &  299 &  13 & &  42 &  45 \\
                         & NGC 4649 & 331 &  39 & & 203 &  10 & &  26 &  17 \\ \cline{2-10}
                         &   total  & 1978 & 298 & & 1254 & 203 & & 254 & 160 \\ \hline
               & NGC 1399 &  75 & 19 & &  45 & 21 & & 14 & 11 \\
               & NGC 4374 & 173 &  0 & & 106 &  0 & & 23 & 10 \\
               & NGC 4472 &  73 &  0 & &  40 &  0 & &  6 &  2 \\
$R_{25} \le R$ & NGC 4486 & 113 & 42 & &  17 & 14 & & 17 &  2 \\
               & NGC 4636 & 171 & 15 & & 148 & 13 & & 27 & 22 \\
               & NGC 4649 & 147 &  6 & &  72 &  6 & & 12 &  6 \\ \cline{2-10}
               &   total  & 752 & 82 & & 428 & 54 & & 99 & 53 \\ \hline \hline
\end{tabular}

\vspace{2truemm}

$^1$ number of sources in optical ground observation only regions

$^2$ number of sources in {\hst} observation regions


\end{minipage}
\end{table}

\clearpage


\begin{table}
\centering
\begin{minipage}{0.85\textwidth}
\caption{Match statistics of LMXBs and GCs\label{tab-psrc2}}
\begin{tabular}{@{}ccccccccc@{}} \hline\hline
Region & Name & LMXB$_{tot}$ & \multicolumn{2}{c}{BGC$-$LMXB} & & \multicolumn{2}{c}{RGC$-$LMXB} & Field$-$LMXB \\ \cline{4-5} \cline{7-8}
       &      &            &     G$^1$ & H$^2$              & &      G$^1$ & H$^2$             & \\ \hline\hline
                      & NGC 1399 & 107 & 5 & 5 & & 28 & 12 & 29 \\
                      & NGC 4374 &  24 & 0 & 3 & &  2 & 2  & 13 \\
                      & NGC 4472 &  76 & 1 & 5 & &  5 & 24 & 33 \\
$20'' \le R < R_{eff}$ & NGC 4486 & 46 & 3 & 4 & &  5 & 9  & 20 \\
                      & NGC 4636 &  23 & 1 & 2 & &  2 & 5  & 13 \\
                      & NGC 4649 &  48 & 0 & 1 & &  7 & 15 & 21 \\ \cline{2-9}
                      &   total  & 324 & 10& 20& & 49 & 67 & 129 \\ \hline
                         & NGC 1399 &  52 & 10 & 1 & & 17 & 0 &  15 \\
                         & NGC 4374 &  33 &  3 & 0 & &  2 & 0 &  18 \\
                         & NGC 4472 &  50 &  6 & 1 & & 13 & 1 &  22 \\
$R_{eff} \le R < R_{25}$ & NGC 4486 &  65 &  9 & 0 & & 16 & 6 &  26 \\
                         & NGC 4636 &  62 &  5 & 0 & & 18 & 0 &  17 \\
                         & NGC 4649 &  79 & 10 & 1 & & 18 & 2 &  31 \\ \cline{2-9}
                         &   total  & 341 & 43 & 3 & & 84 & 9 & 129 \\ \hline
               & NGC 1399 & 18 & 3 & 0 & & 3 & 0 &  8 \\
               & NGC 4374 & 15 & 2 & 0 & & 2 & 0 &  3 \\
               & NGC 4472 &  4 & 0 & 0 & & 1 & 0 &  2 \\
$R_{25} \le R$ & NGC 4486 &  1 & 0 & 0 & & 0 & 0 &  0 \\
               & NGC 4636 & 12 & 1 & 0 & & 2 & 0 &  8 \\
               & NGC 4649 & 12 & 1 & 0 & & 0 & 0 &  9 \\ \cline{2-9}
               &   total  & 62 & 7 & 0 & & 8 & 0 & 30 \\ \hline \hline
\end{tabular}

\vspace{2truemm}

$^1$ number of sources in optical ground observation only regions

$^2$ number of sources in {\hst} observation regions


\end{minipage}
\end{table}



\clearpage


\begin{table}
\footnotesize
\rotate
\centering
\begin{minipage}{1.00\textwidth}
\caption{Summary of match statistics$^1$\label{tab-psrc3}}
\begin{tabular}{@{}cccccccccccccc@{}} \hline\hline
Name & \multicolumn{3}{c}{N(LMXB)} & $N(LMXB_{RGC})\over N(LMXB_{BGC})$ & $N(LMXB_{GC})\over N(LMXB_{Field})$    & \multicolumn{2}{c}{N(OPT)} & $f^2_{BGC}$ & $f^3_{RGC}$ & $f^4_{GC}$ \\ \cline{2-4} \cline{7-8}
         &BGC & RGC & Field &        & &  BGC &  RGC &             &             & \\ \hline\hline
NGC 1399 & 21 &  57 &   44  &          $2.7\pm0.7$   & $1.8\pm0.3$       &  365 &  434 & $5.8\pm1.3$ & $13.1\pm1.9$ & $9.8\pm1.2$ \\
NGC 4374 &  6 &   6 &   31  &          $1.0\pm0.6$   & $0.4\pm0.1$       &  298 &  225 & $2.0\pm0.8$ & $ 2.7\pm1.1$ & $2.3\pm0.7$ \\
NGC 4472 & 13 &  43 &   55  &          $3.3\pm1.0$   & $1.0\pm0.2$       &  610 &  523 & $2.1\pm0.6$ & $ 8.2\pm1.3$ & $4.9\pm0.7$ \\
NGC 4486 & 16 &  36 &   46  &          $2.2\pm0.7$   & $1.1\pm0.2$       &  773 &  740 & $2.1\pm0.5$ & $ 4.9\pm0.8$ & $3.4\pm0.5$ \\
NGC 4636 &  8 &  25 &   30  &          $3.1\pm1.3$   & $1.1\pm0.3$       &  453 &  472 & $1.8\pm0.6$ & $ 5.3\pm1.1$ & $3.6\pm0.6$ \\
NGC 4649 & 12 &  42 &   52  &          $3.5\pm1.1$   & $1.0\pm0.2$       &  510 &  377 & $2.4\pm0.7$ & $11.1\pm1.8$ & $6.1\pm0.9$ \\ \hline
  total  & 76 & 209 &  258  &          $2.8\pm0.4$   & $1.1\pm0.1$       & 2736 & 2771 & $2.8\pm0.3$ & $ 7.5\pm0.5$ & $5.2\pm0.3$ \\ \hline \hline
\end{tabular}
\vspace{2truemm}
$^1$ point sources  for $20'' \le R \le R_{25}$.

$^2$ efficiency to harbor LMXB in \% for blue-GCs, defined by N(LMXB$_{BGC}$)/N(OPT$_{BGC}$)

$^3$ same as $^2$ but for red GCs

$^4$ same as $^2$ but for all GCs
\end{minipage}
\end{table}

\clearpage


\begin{table}
\centering
\begin{minipage}{0.90\textwidth}
\caption{Hardness Ratios and X-ray colors of LMXBs\label{tab-xprop}}
\begin{tabular}{@{}ccccc@{}} \hline\hline
Sample       & Parameter& $R \le R_{eff}/2$ & $R_{eff}/2 < R \le R_{25}/2$ & $R_{25}/2 < R \le R_{25}$ \\ \hline \hline
             & C21      & $-0.23\pm0.32$ & $-0.14\pm0.30$ & $-0.21\pm0.31$ \\
BGC$-$LMXB   & C32      & $+0.85\pm0.51$ & $+0.37\pm0.30$ & $+0.39\pm0.28$ \\
             & HR       & $-0.81\pm0.15$ & $-0.62\pm0.28$ & $-0.73\pm0.25$ \\ \cline{2-5}
             & C21      & $-0.19\pm0.38$ & $-0.18\pm0.27$ & $-0.25\pm0.30$ \\
RGC$-$LMXB   & C32      & $+0.44\pm0.35$ & $+0.45\pm0.33$ & $+0.44\pm0.31$ \\
             & HR       & $-0.65\pm0.44$ & $-0.68\pm0.36$ & $-0.64\pm0.32$ \\ \cline{2-5}
             & C21      & $-0.20\pm0.37$ & $-0.17\pm0.27$ & $-0.24\pm0.31$ \\
 GC$-$LMXB   & C32      & $+0.53\pm0.42$ & $+0.44\pm0.32$ & $+0.42\pm0.30$ \\
             & HR       & $-0.68\pm0.40$ & $-0.67\pm0.35$ & $-0.67\pm0.30$ \\ \cline{2-5}
             & C21      & $-0.14\pm0.44$ & $-0.18\pm0.35$ & $-0.20\pm0.37$ \\
field$-$LMXB & C32      & $+0.59\pm0.44$ & $+0.51\pm0.37$ & $+0.55\pm0.42$ \\
             & HR       & $-0.62\pm0.56$ & $-0.69\pm0.34$ & $-0.71\pm0.35$ \\ \hline \hline
\end{tabular}

\vspace{2truemm}

\end{minipage}
\end{table}

\clearpage


\begin{table}
\centering
\begin{minipage}{0.80\textwidth}
\caption{Linear Least Square fitting of radial profiles$^1$\label{tab-rprof2}}
\begin{tabular}{@{}ccc||ccc@{}} \hline\hline
Sample  & Slope           & $\sigma$   & Sample       & Slope           & $\sigma$   \\ \hline \hline
BGC     & $2.36\pm0.05$  & $0.08$  & BGC$-$LMXB   & $3.71\pm0.54$  & $0.18$  \\
        & ($2.54\pm0.05$)& ($0.07$)&              & ($3.91\pm0.57$)& ($0.20$) \\
RGC     & $3.55\pm0.05$  & $0.08$  & RGC$-$LMXB   & $4.41\pm0.30$  & $0.49$  \\
        & ($3.74\pm0.05$)& ($0.09$)&              & ($4.58\pm0.32$)& ($0.48$) \\
 GC     & $2.87\pm0.03$  & $0.05$  &  GC$-$LMXB   & $4.32\pm0.23$  & $0.18$  \\
        & ($3.06\pm0.03$)& ($0.06$)&              & ($4.50\pm0.24$)& ($0.17$) \\
 Halo   & $4.60\pm0.04$  & $0.05$  & field$-$LMXB & $4.31\pm0.26$  & $0.13$  \\ 
        &                &         & ALL$-$LMXB   & $3.97\pm0.15$  & $0.17$  \\ \hline \hline
\end{tabular}

\vspace{2truemm}

$^1$ Numbers in parenthesis are fitting result for the background-corrected \
GC sample based on Fig. \ref{fig-cont}.

\end{minipage}
\end{table}

\clearpage

\begin{figure}
\epsfig{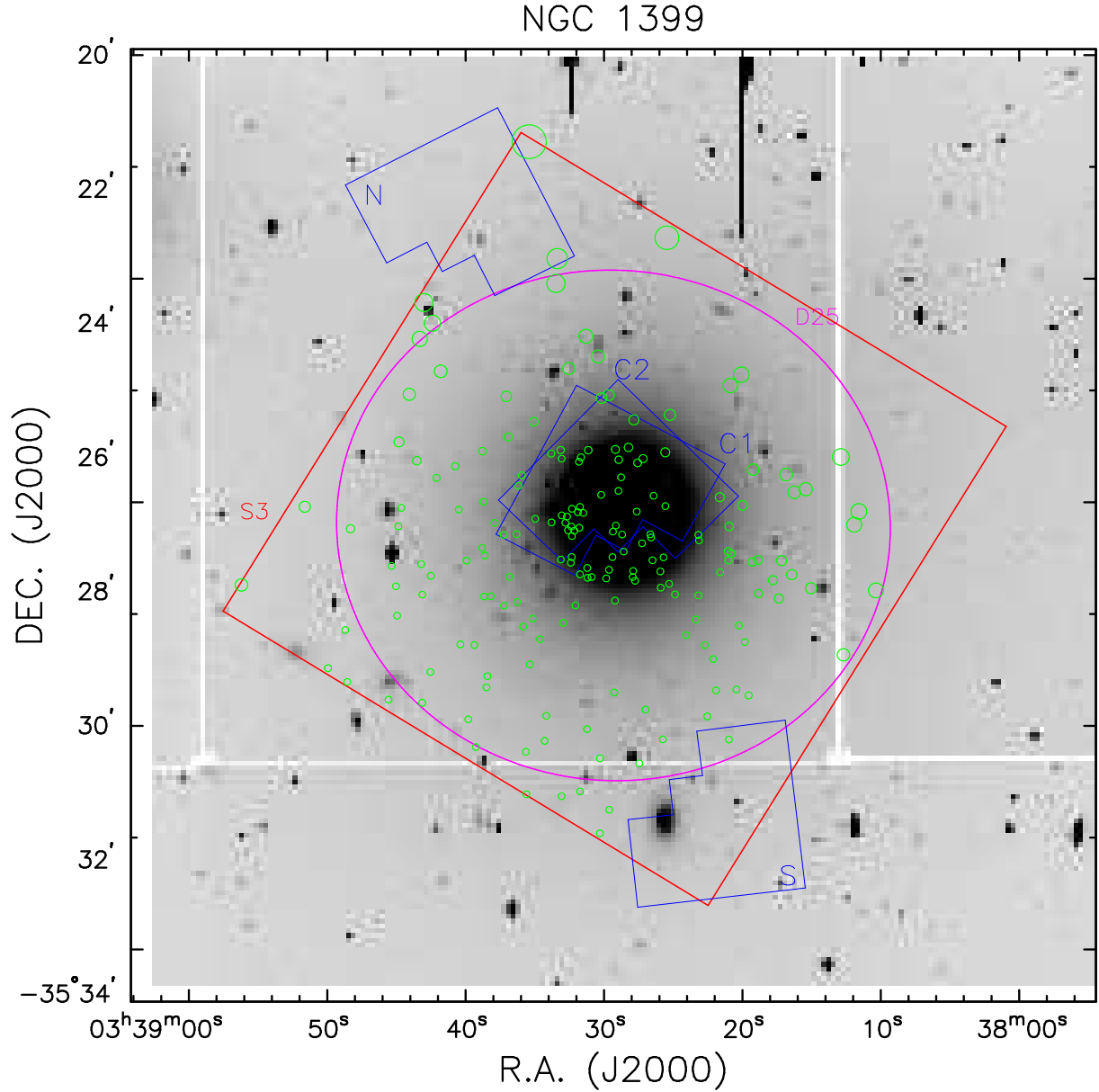}
\vspace{-3mm}
\caption{Observation field of view for NGC 1399. The big square shows the
boundary of the Chandra \s3 chip and the bat-shaped rectangles represent the
{\hstwfpc2} FOVs with field labels (see Table \ref{tab-hobs}). The optical
galaxy is shown with a \D25 ellipse. Point sources detected in \s3 are
shown with small circles.}
\label{fig-fov}
\end{figure}

\clearpage
\begin{center}
\epsfig{figure=a_fig_fov_n4374_small.ps, height=0.975\textwidth, width=0.975\textwidth}
\end{center}
\vspace{-3mm}

\noindent
Fig. \ref{fig-fov} $-$ continued: Observation field of view for NGC 4374.

\clearpage
\begin{center}
\epsfig{figure=a_fig_fov_n4472_small.ps, height=0.975\textwidth, width=0.975\textwidth}
\end{center}
\vspace{-3mm}

\noindent
Fig. \ref{fig-fov} $-$ continued: Observation field of view for NGC 4472.

\clearpage
\begin{center}
\epsfig{figure=a_fig_fov_n4486_small.ps, height=0.975\textwidth, width=0.975\textwidth}
\end{center}
\vspace{-3mm}

\noindent
Fig. \ref{fig-fov} $-$ continued: Observation field of view for NGC 4486

\clearpage
\begin{center}
\epsfig{figure=a_fig_fov_n4636_small.ps, height=0.975\textwidth, width=0.975\textwidth}
\end{center}
\vspace{-3mm}

\noindent
Fig. \ref{fig-fov} $-$ continued: Observation field of view for NGC 4636.

\clearpage
\begin{center}
\epsfig{figure=a_fig_fov_n4649_small.ps, height=0.975\textwidth, width=0.975\textwidth}
\end{center}
\vspace{-3mm}

\noindent
Fig. \ref{fig-fov} $-$ continued: Observation field of view for NGC 4649.

\clearpage

\clearpage

\clearpage

\clearpage

\clearpage

\clearpage

\clearpage

\clearpage

\clearpage

\clearpage

%
%
%

\clearpage

\clearpage

\clearpage

\end{document}